\documentclass[apj]{emulateapj}
\usepackage{color,amsmath,amssymb,latexsym,txfonts,appendix}
\usepackage{graphicx,subfigure,threeparttable,longtable}
\usepackage{array} 

\begin{document}

\title{Studies on cosmic ray nuclei with Voyager, ACE and AMS-02: 
I. local interstellar spectra and solar modulation}
\author{Cheng-Rui Zhu$^{1,2}$, Qiang Yuan$^{1,3,4}$, Da-Ming Wei$^{1,3}$}
\affil{$^1$Key Laboratory of Dark Matter and Space Astronomy, Purple
Mountain Observatory, Chinese Academy of Sciences, Nanjing 210008, China\\
$^2$University of Chinese Academy of Sciences, Beijing 100012, China\\
$^3$School of Astronomy and Space Science, University of Science and
Technology of China, Hefei 230026, Anhui, China\\
$^4$Center for High Energy Physics, Peking University, Beijing 100871, China}
\email{yuanq@pmo.ac.cn (QY), dmwei@pmo.ac.cn (DMW)}

\begin{abstract}
The acceleration of cosmic ray particles and their propagation in the 
Milky Way and the heliosphere tangle with each other, leading to 
complexity and degeneracy of the modeling of Galactic cosmic rays 
(GCRs). The recent measurements of the GCR spectra by Voyager-1 from
outside of the heliosphere gave a first direct observation of GCRs
in the local interstellar (LIS) environment. Together with the 
high-precision data near the Earth taken by ACE and AMS-02, we derive
the LIS spectra of Helium, Lithium, Beryllium, Boron, Carbon, and Oxygen 
nuclei from a few MeV/n to TeV/n, using a non-parameterization method. 
These LIS spectra are helpful in further studying the injection and
propagation parameters of GCRs. The nearly 20 years of data recorded
by ACE are used to determine the solar modulation parameters over the 
solar cycles 23 and 24, based on the force-field approximation. 
We find general agreements of the modulation potential with the 
results inferred from neutron monitors and other cosmic ray data. 
\end{abstract}


\section{Introduction}

It is now widely believed that Galactic cosmic rays (GCRs) get accelerated 
at cosmic accelerators such as shocks of supernova explosions, and then
propagate diffusively in the Galactic random magnetic field. During these 
process, they interact with interstellar gas and fields and then produce 
secondary particles and radiation. After entering the heliosphere, GCRs get 
further modulated by solar winds and the heliospheric magnetic field. 
A detailed modeling of the GCR acceleration and propagation is difficult, 
due to the tanglement of the above mentioned effects and in general only the 
data around the Earth are available. The traditional way is to model each 
of those effects, and fit globally to the data \citep{2009A&A...497..991P,
2010A&A...516A..66P,2011ApJ...729..106T,2015JCAP...09..049J,
2016ApJ...824...16J,2016PhRvD..94l3019K,2016PhRvD..94l3007F,
2017PhRvD..95h3007Y,2018PhRvD..97b3015N}. It is found that there is 
large degeneracy among different models and the corresponding parameters 
\citep{2017PhRvD..95h3007Y,2018PhRvD..97b3015N}.

The Voyager-1 spacecraft, which traveled for about 140 AU from the Earth
in nearly 40 years after its launch in 1977 and crossed the boundary of
the heliosphere on August 25, 2012 \citep{2013Sci...341..150S}, provides
us a unique opportunity to observe GCRs in the local interstellar space 
(LIS) for the first time. Together with the measurements on top of the
atmosphere (TOA) near the Earth, the Voyager-1 data can provide very
useful constraints on the source and propagation parameters of GCRs,
as well as the solar modulation effect \citep{2016ApJ...831...18C,
2016PhRvD..93d3016C,2016A&A...591A..94G,2016ApJ...829....8C,
2017ApJ...840..115B}. 

Recently, the AMS-02 collaboration reported the measurements of energy
spectra of primary nuclei (He, C, and O) and secondary ones (Li, Be, 
and B) to rigidities of a few TV with very high precision 
\citep{2017PhRvL.119y1101A,2018PhRvL.120b1101A}.
The AMS-02 data showed similar properties of the primary nuclei
or secondary ones at high energies. Interestingly, it has been found
that the spectra of all these particles experienced hardenings at
a few hundred GV, and the secondary family hardens even more than
that of the primary family which has very important implication in
understanding the physics of the spectral hardenings
\citep{2017PhRvL.119x1101G,2018PhRvD..97f3008G,2018arXiv180203602L}.

The data of Voyager-1 and AMS-02 together give a full description 
of spectral behaviors of various nuclei from MeV to TeV energies.
In particular, the inclusion of the Voyager-1 data may help  
to effectively break the degeneracy between the LIS spectra and the
solar modulation effect. In this paper and a forthcoming one 
(Zhu et al., in preparation), we will study the injection and 
propagation properties of GCRs based on these new observational data. 
In addition to the Voyager-1 and AMS-02 data,
the long-term monitored data by the Cosmic Ray Isotope Spectrometer 
(CRIS) on the Advanced Composition Explorer (ACE) spacecraft are
also employed to provide further constraints on the GCR spectra 
in the gap region between Voyager-1 and AMS-02. In this paper we 
derive the LIS spectra of GCR nuclei from He to O, using a 
non-parameteric method \citep{2016A&A...591A..94G}. The time
series of the solar modulation parameter over the past 20 years
will also be studied according to the ACE-CRIS data. The propagation
parameters of GCRs will be investigated in detail in the next paper.

\section{Methodology}

\subsection{Solar modulation}
GCRs would get modulated by the heliospheric magnetic field carried by
solar winds when they enter the heliosphere, resulting in suppression
of their fluxes. This solar modulation effect depends on particle 
energies, and is particularly obvious at low energies. In this work
we adopt the force-field approximation of the solar modulation
\citep{1967ApJ...149L.115G,1968ApJ...154.1011G}, which was actually an 
approximate solution of the Parker's equation \citep{1965P&SS...13....9P}. 
In this model, the TOA flux is related with the LIS flux as
\begin{equation}\label{force_filed}
J^{\rm TOA}(E)=J^{\rm LIS}(E+\Phi)\times\frac{E(E+2m_p)}
{(E+\Phi)(E+\Phi+2m_p)}, 
\end{equation}
where $E$ is the kinetic energy per nucleon, $\Phi=\phi\cdot Z/A$ with 
$\phi$ being the solar modulation potential, $m_p=0.938$ GeV is the 
proton mass, and $J$ is the differential flux of GCRs. The only
parameter in the force-field model is the modulation potential $\phi$.
In principle, the force-field model assumes a quasi-steady-state of 
the solution of the Parker's equation. However, the observational GCR 
fluxes show 11-year variations associated with solar activities.
Therefore a time-series of $\phi$ at different epochs is adopted to
describe the data.

\subsection{Non-parametric flux: splines}

Usually power-law or broken power-law functions are exployed to fit the 
GCR data. If the observational data cover a wide enough energy range, one 
can instead use a non-parametric method by means of spline interpolation 
of GCR fluxes among a few knots \citep{2016A&A...591A..94G}. 
The spline interpolation is a way to obtain an approximate function 
smoothly passing through a series of points using piecewise polynomial 
functions. We use the cubic spline interpolation here, with the 
highest-order of polynomial of three. We work in the $\log(E)-\log(J)$ 
space of the energy spectrum. The positions of knots of $x=\log(E)$ for
Helium, Boron, Carbon, and Oxygen are defined as
\begin{equation}
\begin{split}
    &\{x_1,x_2,x_3,x_4,x_5,x_6,x_7,x_8,x_9\} \\
    =&\{-2.3,-1.6,-0.9,-0.2,0.5,1.2,1.9,2.6,3.3\}.
\end{split}
\end{equation}
For Lithium and Beryllium nuclei, the numbers of Voyager-1 data points 
are very limited, and their number of knots are adopted to be 7, as
\begin{equation}
\begin{split}
    &[x_1,x_2,x_3,x_4,x_5,x_6,x_7] \\
    =&\{-1.6,-0.83,-0.06,0.71,1.48,2.25,3.3\}.
\end{split}
\end{equation}
We also check the results through adding or reducing the number of knots, 
and find that the results change only slightly in the energy region where 
no data are available. In the following, $y_i$ parameters at the
above fixed $x_i$ knot positions are assumed to be free and are derived 
through fitting to the data.


\subsection{Data sets}

The GCR data from AMS-02 \citep{2017PhRvL.119y1101A,2018PhRvL.120b1101A}, 
Voyager-1 \citep{2016ApJ...831...18C}, and 
ACE-CRIS\footnote{\url{http://www.srl.caltech.edu/ACE/ASC/level2/lvl2DATA\_CRIS.html}}
are adopted. For AMS-02 and Voyager-1, the data about Helium, Lithium, 
Beryllium, Boron, Carbon, Oxygen nuclei are available, while for ACE-CRIS 
only the Boron, Carbon, and Oxygen data are available. The AMS-02 data 
were taken between May 19, 2011 to May 26, 2016. We extract the ACE-CRIS 
data of the same period from the ACE Science center to derive the LIS 
spectra. The ACE data of the whole 20 years of operation are then used 
to study the solar modulation. The uncertainties of the ACE data are the 
quadratical sum of the statistical ones and the systematic ones, with 
the latter mainly coming from the geometry factor ($2\%$), the 
scintillating optical fiber trajectory efficiency ($2\%$), and the 
spallation correction ($1\% \sim 5\%$) \citep{2009ApJ...698.1666G}. 
Note that, the proton spectra by AMS-02 \citep{2015PhRvL.114q1103A} 
and Voyager-1 \citep{2016ApJ...831...18C} are not included in this work. 
This is because the data-taking time for protons of AMS-02 is different 
from the other nuclei, which may complicate the solar modulation modeling 
when fitting the LIS spectra. Furthermore, protons are less relevant in
the study of GCR propagation compared with the primary and secondary
nuclei discussed in this work.

\subsection{$\chi^2$ analysis}

We fit the normalizations of the $n$ spline knots, together with the
solar modulation potential $\phi$. The $\chi^2$ statistics is defined as
\begin{eqnarray}
\chi^2=\sum_{i=1}^{m}\frac{{\left[J(E_i;\boldsymbol{y},\phi)-
J_i(E_i)\right]}^2}{{\sigma_i}^2},
\end{eqnarray}
where $J(E_i;\boldsymbol{y},\phi)$ is the expected flux, $J_i(E_i)$ and
$\sigma_i$ are the measured flux and error for the $i$th data bin with
central energy $E_i$. 

We use the Markov Chain Monte Carlo (MCMC) algorithm to minimize the
$\chi^2$ function, which works in the Bayesian framework. The posterior
probability of model parameters $\boldsymbol{\theta}$ is given by
\begin{equation}
p(\boldsymbol{\theta}|{\rm data}) \propto {\mathcal L}(\boldsymbol{\theta})
p(\boldsymbol{\theta}),
\end{equation}
where ${\mathcal L}(\boldsymbol{\theta})$ is the likelihood function
of parameters $\boldsymbol{\theta}$ given the observational data, and 
$p(\boldsymbol{\theta})$ is the prior probability of $\boldsymbol{\theta}$.

The MCMC driver is adapted from {\tt CosmoMC} \citep{2002PhRvD..66j3511L}.
We adopt the Metropolis-Hastings algorithm. The basic procedure of this
algorithm is as follows. We start with a random initial point in the 
parameter space, and jump to a new one following the covariance of these
parameters. The accept probability of this new point is defined as
$\min\left[p(\boldsymbol{\theta}_{\rm new}|{\rm data})/p(\boldsymbol{\theta}_
{\rm old}|{\rm data}),1\right]$. If the new point is accepted, then repeat
this procedure from this new one. Otherwise go back to the old point.
For more details about the MCMC one can refer to \citep{MCMC}.

\section{Results}

\subsection{LIS fluxes of various nuclei}

\begin{figure}[ht!]
    \centering
    \includegraphics[scale=0.35]{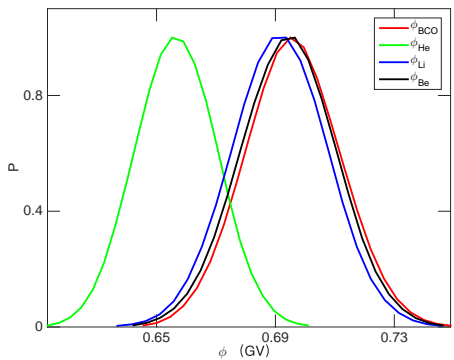}
    \caption{The probability density distributions of $\phi_{\rm BCO}$, 
    $\phi_{\rm He}$, $\phi_{\rm Li}$ and $\phi_{\rm Be}$. All the curves
    are normalized to a peak value of unit.}
    \label{fig:pdf}
\end{figure}

The solar modulation degenerates with the LIS fluxes. To constrain the
solar modulation potential as effectively as possible, we jointly fit the 
Boron, Carbon, and Oxygen data, for which the low energy measurements
from both Voyager and ACE data are available. This fits gives 
$\phi_{\rm BCO}=0.696\pm0.016$ GV. Using this value as a prior, we then
fit the Helium, Lithium, and Beryllium data, and get $\phi_{\rm He}=0.657 
\pm 0.013$ GV, $\phi_{\rm Li}=0.692 \pm 0.016$ GV, $\phi_{\rm Be}=0.694 
\pm 0.016$ GV. We find that all these fits give $\phi\sim0.7$ GV for the 
average solar modulation potentials between May 19, 2011 to May 26, 2016, 
except for He which gives a somehow smaller modulation potential. 
Table~\ref{tab:chi2} gives the best-fitting $\chi^2$ values and the 
modulation potentials (with $1\sigma$ uncertainties). 
The probability distribution functions of $\phi_{\rm BCO}$, 
$\phi_{\rm He}$, $\phi_{\rm Li}$, and $\phi_{\rm Be}$ are shown in
Fig.~\ref{fig:pdf}.

\begin{figure}[ht!]
    \centering
    \includegraphics[scale=0.35]{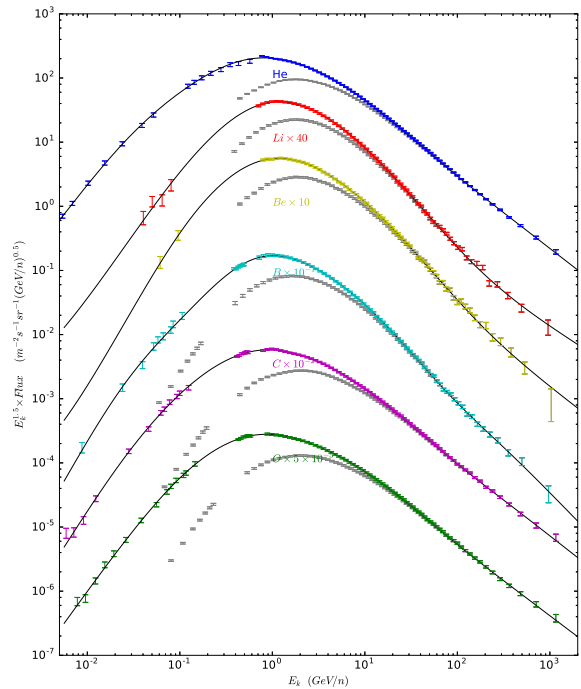}
    \caption{Best-fit LIS fluxes (lines), multiplied by 
    $E_k^{1.5}$, compared with the measurements (colorful points) 
    of Voyager-1 \citep{2016ApJ...831...18C}, and the de-modulated 
    results of AMS-02 and ACE. The TOA measurements of AMS-02 
    \citep{2017PhRvL.119y1101A,2018PhRvL.120b1101A} and ACE are shown by 
    gray points.}
    \label{fig:flux}
\end{figure}

The best-fit LIS spectra of all these nuclei are shown by solid lines 
in Fig.~\ref{fig:flux}. We can see that this non-parametric method 
reproduces reasonably any 
broad structures of the energy spectrum, such as the breaks at $O(1)$ 
and $O(100)$ GeV/n. We use the fitted results of $\phi_{\rm He}$, 
$\phi_{\rm Li}$, $\phi_{\rm Be}$, and $\phi_{\rm BCO}$ to de-modulate 
the TOA measurements by ACE and AMS-02 to obtain the corresponding LIS 
fluxes, as shown by the colored data points in Fig.~\ref{fig:flux}. 
The uncertainties associated with the modulation parameter, obtained 
using the error propagation, are added quadratically to the original 
(statistical and systematic) uncertainties of the measurements. 
For kinetic energies smaller than $\sim 1$ GeV/n, the 
uncertainties due to the modulation parameter account for $\sim10\%$ 
of the total ones, which become smaller at higher energies.
The results of the LIS fluxes are given in Tables~\ref{tab:He}$-$\ref{tab:O} 
in the Appendix.
 
In Fig.~\ref{fig:flux_uni}, we compare the fitted $2\sigma$ results of
the LIS fluxes for the primary group (He, C, O) and secondary group
(Li, Be, B), with proper normalizations. For the primary group, the energy 
spectra of He, C, and O are similar with each other for energies above 
$\sim1$ GeV/n. The low energy spectrum of He is different from
that of C and O, which is possibly due to different energy loss rates of 
them in the interstellar medium. Whether there are differences among the 
injection spectra of these primary nuclei needs detailed studies within
specified propagation models. The spectra of secondary nuclei are in
agreement with each other within the uncertainties.

\begin{table}[ht!]
\caption{Solar modulation potentials $\phi$ and $\chi^2$ values of various 
fits.}
\centering
\begin{tabular}{cccccc}
    \hline\hline
    Species & $\phi$ (GV) & $\chi^2/dof$ \\
    \hline
    Helium  & $0.657\pm0.013$ & 48.2/85 \\
    Lithium  & $0.692\pm0.016$ & 39.4/72 \\
    Beryllium  & $0.694\pm0.016$ & 29.4/70 \\
    BCO & $0.696\pm0.016$ & 81.9/258 \\
    \hline
\end{tabular}
\label{tab:chi2}
\end{table} 
  
\begin{figure}[ht!]
    \centering
    \includegraphics[scale=0.35]{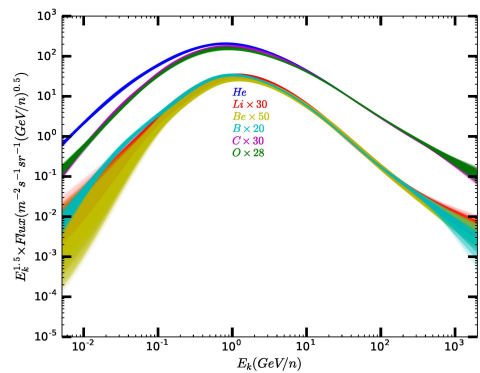}
    \caption{The $2\sigma$ bands of the LIS fluxes, multiplied 
    by $E_k^{1.5}$, of different nuclei.}  
    \label{fig:flux_uni}
\end{figure}

\subsection{Time series of $\phi$}

Given the LIS fluxes of CRs, we can then obtain the time evolution of
the solar modulation potentials using the long-term measurements of ACE.
The ACE data in each Bartels rotation period (27 days) from 1997 to 2016 
are extracted. Using the LIS spectra of Boron, Carbon, and Oxygen nuclei, 
we can derive monthly values of the solar modulation potential. 
A Bayesian approach is adopted to properly take into account the 
uncertainties of the LIS spectra. The posterior probability of $\phi$ 
is given by
\begin{equation}
p(\phi)\propto\int {\mathcal L}_{\rm ACE}(\phi,\boldsymbol{y})
p(\boldsymbol{y}){\rm d}\boldsymbol{y},
\end{equation}
where ${\mathcal L}_{\rm ACE}\propto \exp\left(-\chi^2_{\rm ACE}/2\right)$ 
is the likelihood of model parameters ($\phi,\boldsymbol{y}$), 
$p(\boldsymbol{y})$ is the prior probability distribution of 
$\boldsymbol{y}$ which is obtained through the fit in Sec.~3.1. 
The above integration is simply calculated through adding the parameter 
sets of the last 50\% of the Markov chains together, weighted by their
stopping numbers at each point.

\begin{figure}[ht!]
    \centering
    \includegraphics[scale=0.4]{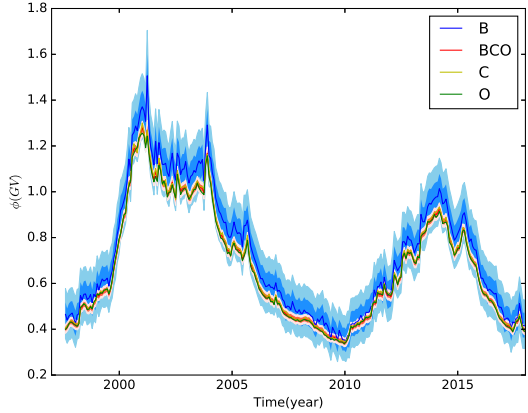} 
    \caption{Time series of $\phi$ via fitting to the ACE Boron, Carbon, 
    Oxygen data from 1997 to 2017.} 
    \label{fig:BCO}
\end{figure}

The posterior mean values (solid lines) and the associated $1\sigma$ and
$2\sigma$ bands (thick and thin shaded regions) of $\phi$ for the fittings 
to Boron, Carbon, Oxygen nuclei individually and simultaneously are shown 
in Fig.~\ref{fig:BCO}. We find that the Carbon and Oxygen data give very 
close results of the modulation potential, while the Boron data give 
slightly larger results. Since the fluxes of Boron are lower than that 
of Carbon and Oxygen, the corresponding uncertainties of $\phi$ derived 
from the Boron data are also larger. Within the uncertainties, these 
results are consistent with each other. We tabulate the 27-day
time series of $\phi_{\rm BCO}$ and the associated lower and upper limits
in Table \ref{tab:phi}.

\begin{figure}[ht!]
    \centering
    \includegraphics[scale=0.4]{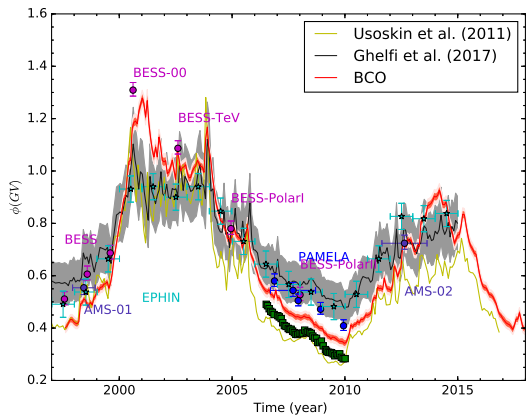}
    \caption{Time series of the modulation potential $\phi$. The red line
    and associated bands are derived in this work via the joint fit to
    Boron, Carbon, and Oxygen data. Previous results inferred from the
    neutron monitor data \citep{2011JGRA..116.2104U,2017AdSpR..60..833G}
    and GCR data \citep{2016ApJ...829....8C,2016A&A...591A..94G,
    2017AdSpR..60..833G} are also shown for comparison.}
    \label{fig:phi}
\end{figure}

Fig.~\ref{fig:phi} compares our results (red curve and associated 68\% and
95\% bands) of the modulation potential for the joint fit with previous 
results. The gray line and shaded band show the monthly results\footnote{\url{http://lpsc.in2p3.fr/crdb/}} 
from netron monitors given in \citet{2017AdSpR..60..833G}, and the yellow
line represents also the 
results\footnote{\url{http://cosmicrays.oulu.fi/phi/Phi\_mon.txt}} derived 
from neutron monitor data given in \citet{2011JGRA..116.2104U}. Other data 
points are derived from the studies of various GCR data 
\citep{2016ApJ...829....8C,2016A&A...591A..94G,2017AdSpR..60..833G}.
The results from different analyses show rough consistency with each other.
Quantitatively, they may differ by as large as $50\%$, in particular for
the periods of solar maximum around 2001 or minimum around 2010. 
The difference may come from different energy ranges of
relevant data sets, and/or assumptions of the LIS spectra of GCRs 
adopted in different works. One improvement in our work is the use of
the Voyager-1 data taken outside of the solar system to constrain the 
LIS spectra of GCR nuclei, which makes our LIS spectra less uncertain
compared with most of previous studies. 
 
\begin{figure}[ht!]
    \centering
    \includegraphics[scale=0.4]{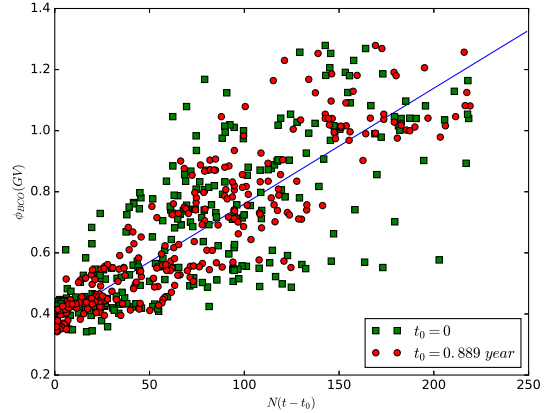}
    \caption{The scatter plot between $\phi_{\rm BCO}(t)$ and 
    sunspot number $N$ without (squares) and with (dots) time delay $t_0$. 
    The solid line shows a linear fit using Eq.~(\ref{sunspots_phi_l}) 
    with $t_0=0.889$~yr.}
    \label{fig:ss_phi}
\end{figure} 
    
Another indicator of solar activities is the sunspot number. Observational 
evidence shows a strong correlation between the sunspot numbers and solar 
activities. Fig.~\ref{fig:ss_phi} shows the relationship between the
solar modulation potential $\phi_{\rm BCO}(t)$ obtained in this work
and the sunspot numbers\footnote{\url{http://sidc.oma.be/silso/datafiles}} 
at time $t-t_0$, where $t_0$ represents a time delay from the solar activity 
to the modulation of GCRs. We assume a linear correlation between them, as
\begin{equation}
\label{sunspots_phi_l}
\phi(t)=\phi_1+\phi_2\times \frac{N(t-t_0)}{100},
\end{equation}
which is shown by the solid line in Fig.~\ref{fig:ss_phi}. The fitting 
parameters are given in Table~\ref{tab:ss}. Note that the sunspot numbers 
fluctuate significantly, and thus the uncertainties of the parameters are 
statistically meaningless. The fit gives a time delay of $\sim0.9$ 
yr, which can be understood as the time for solar winds traveling across 
the solar system ($\sim100$ astronomical units) with a typical speed of 
$\sim500$ km s$^{-1}$ \citep{2017PhRvD..95h3007Y}. The results 
without time delay are also shown in Fig.~\ref{fig:ss_phi} for comparison. 
We can see that the scattering of data points are clearly larger in case 
of no time delay. Similar time delay was also found in previous works 
\citep[e.g.,][]{2017ApJ...849L..32T,2017JGRA..122.1463K}. 
\citet{2017ApJ...849L..32T} found a time delay of $0.68\pm0.10$~yr, 
which is consistent with ours within $2\sigma$ level. Different time 
delays in the even ($\sim0.5$~yr) and odd ($\sim1.3$~yr) solar cycles 
were suggested in \citet{2017JGRA..122.1463K}, whose average is fairly 
consistent with our result.

\begin{table}[ht!]
    \caption{The parameters of the linear correlation between $\phi$ and $N$.}
    \centering
    \begin{tabular}{ccc}
    \hline\hline
    Parameters & Unit & Value  \\
    \hline
    $\phi_1$ & GV & $0.378\pm 0.001$\\
    $\phi_2$ & GV & $0.381\pm 0.001$\\
    $t_0$    & yr & $0.889\pm 0.001$\\
    \hline
    \end{tabular}
    \label{tab:ss}
\end{table}
 
\section{Conclusion and discussion}

In this work we use the recent measurements of the GCR fluxes of several
nuclei in an energy range from several MeV/n to TeV/n by Voyager-1, 
ACE-CRIS, and AMS-02 to derive the LIS spectra of GCRs by means of a 
non-parametric spline interpolation method. Through fitting to the data 
of Helium, Lithium, Beryllium, Boron, Carbon, and Oxygen nuclei, we 
obtain very similar solar modulation parameters for different nuclei.
Based on this result, we de-modulate the ACE and AMS-02 observations
from the TOA to the LIS, which can be used in further studies of the
injection and propagation of GCRs. We further derive the time series
of the solar modulation potential according to the 20 years of ACE
measurements of Boron, Carbon, and Oxygen data. Our results of the
solar modulation potential are fairly consistent with previous works. 

The solar modulation parameters differ up to several tens 
percents among different works based on different methods/data. 
Several kinds of reasons may result in such differences. The 
differences of the $\phi$ series between ours and that derived in 
\citet{2011JGRA..116.2104U} using the data from neutron monitors 
are probably due partly to different assumptions of the LIS spectra.
Both \citet{2017AdSpR..60..833G} and this work use a similar
non-parametric method as well as the Voyager-1 data to obtain the 
LIS spectra. However, our results are systematically smaller than theirs 
during the solar minimum, and larger instead during the solar maximum.
Such differences may be due to different fitting energy ranges in these 
works. The data from ACE-CRIS range from $\sim$50 to 200 MeV/n, while 
neutron monitor data are more sensitive to cosmic rays with energies
$\gtrsim$10 GeV/n. Different energy ranges of data may lead to 
systematically different results of the modulation
\citep{2017JGRA..12210964G,2017PhRvD..96j3005T}. This may also explain
the differences between our results and that derived based on other GCR
data which are mainly available at higher energies than ACE-CRIS.
The dependence on the analyzed energy range may reflect the limitation 
of the force-filed approximation in describing the GCR modulation in a 
very wide energy range \citep{2016ApJ...829....8C,2017JGRA..12210964G}.
Finally, there may also be uncertainties in the modeling of neutron
yields in the atmosphere.

As discussed above, the simple force-field model may not be 
precise enough to describe the wide-band GCR modulation. When the
polarity of the solar magnetic field changes at solar maximum, the
force-field model may also fail due to the non-realistic assumption
of the modulation process. The extension of the current work with 
more realistic modulation models, such as that discussed in 
\citet{2016CoPhC.207..386K}, should be important and will be explored
in future works.


\acknowledgments
We thank the ACE-CRIS instrument team and the ACE Science Center for
providing the ACE data. This work is supported by the National Key
Research and Development Program of China (No. 2016YFA0400204), 
the Key Research Program of Frontier Sciences of Chinese Academy of 
Sciences (No. QYZDJ-SSW-SYS024), the National Natural Science Foundation 
of China (Nos. 11722328, U1738205, 11433009), and the 100 Talents program 
of Chinese Academy of Sciences.

\bibliographystyle{aasjournal}
\bibliography{refs}

\clearpage

\appendix

\begin{longtable}{cccccc||cccccc}

    \caption{Time series of $\phi_{\rm BCO}$ and the $1\sigma$ and $2\sigma$
    lower (upper) limits.}\\
    
    \setlength{\extrarowheight}{2pt}\\
    \toprule
    $t$~(year) & $\phi$~(GV) & LO~(1$\sigma$) & UP~(1$\sigma$) & LO~(2$\sigma$) & UP~(2$\sigma$) & $t$~(year) & $\phi$~(GV) & LO~(1$\sigma$) & UP~(1$\sigma$) & LO~(2$\sigma$) & UP~(2$\sigma$)\\
    \hline

    \endfirsthead
    \caption[]{(continued)}\\
    \toprule
    year&$\phi$(GV)&lower1&upper1&lower2&uppper2&year&$\phi$(GV)&lower1&upper1&lower2&uppper2\\
    \hline

    \endhead
    \hline
    \endfoot
    1997.616&0.402&0.391&0.411&0.382&0.420&2007.824&0.442&0.431&0.453&0.422&0.462\\
    1997.690&0.409&0.399&0.421&0.390&0.430&2007.898&0.444&0.433&0.454&0.424&0.464\\
    1997.764&0.423&0.413&0.431&0.403&0.440&2007.972&0.442&0.431&0.453&0.422&0.461\\
    1997.838&0.433&0.422&0.443&0.413&0.452&2008.046&0.438&0.427&0.450&0.418&0.459\\
    1997.912&0.432&0.421&0.444&0.412&0.452&2008.120&0.441&0.431&0.451&0.421&0.461\\
    1997.986&0.424&0.413&0.436&0.404&0.444&2008.194&0.447&0.434&0.455&0.425&0.464\\
    1998.060&0.419&0.408&0.430&0.398&0.439&2008.268&0.441&0.433&0.456&0.424&0.465\\
    1998.134&0.417&0.406&0.428&0.397&0.436&2008.342&0.440&0.429&0.451&0.420&0.460\\
    1998.208&0.424&0.413&0.433&0.404&0.442&2008.416&0.440&0.429&0.453&0.421&0.461\\
    1998.282&0.492&0.481&0.503&0.473&0.512&2008.490&0.433&0.422&0.443&0.412&0.452\\
    1998.356&0.497&0.485&0.507&0.476&0.517&2008.564&0.426&0.415&0.437&0.406&0.445\\
    1998.430&0.511&0.499&0.520&0.490&0.530&2008.638&0.413&0.401&0.422&0.392&0.430\\
    1998.504&0.507&0.495&0.516&0.485&0.526&2008.712&0.413&0.401&0.428&0.393&0.434\\
    1998.578&0.490&0.478&0.501&0.469&0.510&2008.786&0.407&0.397&0.418&0.388&0.426\\
    1998.652&0.488&0.476&0.496&0.466&0.506&2008.860&0.400&0.390&0.409&0.381&0.418\\
    1998.726&0.498&0.486&0.508&0.477&0.517&2008.934&0.399&0.387&0.409&0.379&0.416\\
    1998.800&0.488&0.476&0.498&0.467&0.508&2009.008&0.398&0.387&0.407&0.378&0.415\\
    1998.874&0.515&0.503&0.524&0.493&0.533&2009.082&0.383&0.372&0.393&0.364&0.400\\
    1998.948&0.522&0.511&0.531&0.500&0.540&2009.156&0.380&0.369&0.389&0.360&0.397\\
    1999.021&0.550&0.538&0.560&0.528&0.570&2009.230&0.370&0.359&0.379&0.351&0.387\\
    1999.095&0.545&0.533&0.557&0.524&0.566&2009.304&0.367&0.357&0.378&0.348&0.385\\
    1999.169&0.556&0.545&0.569&0.536&0.579&2009.378&0.359&0.352&0.375&0.343&0.380\\
    1999.243&0.557&0.545&0.568&0.536&0.578&2009.452&0.365&0.355&0.375&0.346&0.382\\
    1999.317&0.568&0.555&0.578&0.545&0.588&2009.526&0.365&0.352&0.372&0.341&0.377\\
    1999.391&0.570&0.558&0.580&0.547&0.590&2009.600&0.360&0.349&0.371&0.341&0.378\\
    1999.465&0.577&0.565&0.587&0.554&0.597&2009.674&0.354&0.346&0.367&0.336&0.373\\
    1999.539&0.552&0.541&0.563&0.531&0.573&2009.748&0.350&0.340&0.359&0.332&0.366\\
    1999.613&0.573&0.562&0.584&0.552&0.594&2009.821&0.349&0.339&0.361&0.331&0.368\\
    1999.687&0.608&0.596&0.620&0.587&0.629&2009.895&0.348&0.338&0.361&0.330&0.367\\
    1999.761&0.662&0.650&0.675&0.641&0.686&2009.969&0.342&0.331&0.354&0.324&0.361\\
    1999.835&0.702&0.689&0.714&0.679&0.725&2010.043&0.345&0.334&0.354&0.325&0.363\\
    1999.909&0.741&0.728&0.754&0.718&0.765&2010.117&0.358&0.348&0.370&0.339&0.377\\
    1999.983&0.783&0.770&0.796&0.760&0.807&2010.191&0.378&0.366&0.387&0.358&0.398\\
    2000.057&0.817&0.804&0.829&0.792&0.841&2010.265&0.412&0.400&0.421&0.391&0.431\\
    2000.131&0.856&0.841&0.868&0.830&0.880&2010.339&0.403&0.394&0.415&0.386&0.424\\
    2000.205&0.893&0.879&0.906&0.868&0.919&2010.413&0.414&0.404&0.422&0.394&0.432\\
    2000.279&0.902&0.889&0.915&0.877&0.927&2010.487&0.415&0.404&0.424&0.394&0.433\\
    2000.353&0.979&0.964&0.992&0.953&1.005&2010.561&0.425&0.413&0.435&0.404&0.445\\
    2000.427&1.037&1.021&1.050&1.007&1.063&2010.635&0.420&0.408&0.429&0.399&0.439\\
    2000.501&1.040&1.025&1.055&1.011&1.069&2010.709&0.433&0.421&0.441&0.412&0.452\\
    2000.575&1.164&1.149&1.179&1.135&1.193&2010.783&0.432&0.421&0.441&0.411&0.451\\
    2000.649&1.186&1.170&1.201&1.156&1.216&2010.857&0.438&0.427&0.448&0.417&0.458\\
    2000.723&1.191&1.175&1.205&1.161&1.219&2010.931&0.451&0.440&0.461&0.431&0.471\\
    2000.797&1.181&1.165&1.196&1.151&1.211&2011.005&0.449&0.437&0.459&0.428&0.468\\
    2000.871&1.230&1.214&1.246&1.199&1.262&2011.079&0.453&0.442&0.462&0.432&0.471\\
    2000.945&1.253&1.238&1.268&1.223&1.283&2011.153&0.456&0.447&0.468&0.440&0.479\\
    2001.019&1.279&1.262&1.294&1.247&1.309&2011.227&0.503&0.491&0.512&0.481&0.522\\
    2001.093&1.257&1.242&1.272&1.227&1.287&2011.301&0.518&0.506&0.528&0.497&0.537\\
    2001.167&1.206&1.190&1.220&1.176&1.234&2011.375&0.513&0.502&0.525&0.494&0.535\\
    2001.241&1.269&1.248&1.290&1.228&1.311&2011.449&0.556&0.545&0.568&0.536&0.579\\
    2001.315&1.180&1.164&1.194&1.149&1.209&2011.523&0.551&0.540&0.563&0.531&0.573\\
    2001.389&1.125&1.110&1.140&1.096&1.154&2011.597&0.552&0.542&0.565&0.534&0.576\\
    2001.463&1.081&1.066&1.095&1.052&1.108&2011.671&0.547&0.535&0.558&0.525&0.568\\
    2001.537&1.054&1.039&1.068&1.026&1.081&2011.745&0.591&0.580&0.604&0.571&0.615\\
    2001.611&1.091&1.076&1.105&1.061&1.119&2011.819&0.562&0.550&0.573&0.541&0.583\\
    2001.684&1.054&1.039&1.069&1.025&1.083&2011.893&0.548&0.538&0.561&0.528&0.572\\
    2001.758&1.129&1.114&1.145&1.099&1.159&2011.967&0.550&0.538&0.560&0.527&0.571\\
    2001.832&1.097&1.082&1.113&1.067&1.128&2012.041&0.543&0.532&0.554&0.522&0.565\\
    2001.906&1.040&1.025&1.054&1.012&1.067&2012.115&0.565&0.553&0.576&0.544&0.586\\
    2001.980&1.035&1.018&1.050&1.004&1.066&2012.189&0.637&0.624&0.649&0.614&0.660\\
    2002.054&1.040&1.025&1.053&1.011&1.066&2012.263&0.623&0.611&0.635&0.601&0.645\\
    2002.128&0.990&0.976&1.003&0.962&1.016&2012.337&0.596&0.584&0.606&0.573&0.617\\
    2002.202&0.996&0.982&1.010&0.969&1.023&2012.411&0.617&0.605&0.630&0.596&0.640\\
    2002.276&1.016&1.000&1.029&0.986&1.043&2012.484&0.627&0.615&0.639&0.605&0.650\\
    2002.350&1.041&1.027&1.054&1.013&1.067&2012.558&0.714&0.701&0.726&0.691&0.738\\
    2002.424&1.005&0.990&1.018&0.977&1.032&2012.632&0.732&0.719&0.746&0.709&0.757\\
    2002.498&0.968&0.954&0.982&0.941&0.996&2012.706&0.710&0.697&0.722&0.686&0.733\\
    2002.572&1.082&1.067&1.096&1.053&1.110&2012.780&0.740&0.727&0.752&0.716&0.764\\
    2002.646&1.046&1.030&1.059&1.017&1.073&2012.854&0.731&0.718&0.744&0.708&0.755\\
    2002.720&1.002&0.987&1.015&0.974&1.028&2012.928&0.737&0.724&0.750&0.714&0.761\\
    2002.794&1.012&0.997&1.026&0.984&1.039&2013.002&0.715&0.703&0.728&0.692&0.739\\
    2002.868&1.016&1.002&1.030&0.989&1.043&2013.076&0.692&0.680&0.705&0.669&0.716\\
    2002.942&1.024&1.009&1.037&0.996&1.050&2013.150&0.691&0.678&0.703&0.668&0.714\\
    2003.016&0.992&0.977&1.006&0.965&1.019&2013.224&0.717&0.704&0.728&0.693&0.740\\
    2003.090&0.974&0.959&0.987&0.946&1.000&2013.298&0.712&0.699&0.724&0.688&0.735\\
    2003.164&0.976&0.961&0.988&0.949&1.001&2013.372&0.806&0.792&0.819&0.780&0.831\\
    2003.238&1.000&0.985&1.013&0.972&1.026&2013.446&0.814&0.801&0.828&0.790&0.840\\
    2003.312&1.011&0.996&1.024&0.984&1.038&2013.520&0.820&0.806&0.833&0.795&0.845\\
    2003.386&1.019&1.004&1.033&0.992&1.047&2013.594&0.832&0.818&0.845&0.807&0.857\\
    2003.460&1.041&1.025&1.054&1.012&1.067&2013.668&0.868&0.854&0.880&0.842&0.892\\
    2003.534&1.028&1.013&1.040&0.999&1.054&2013.742&0.871&0.857&0.884&0.845&0.896\\
    2003.608&1.018&1.004&1.031&0.989&1.044&2013.816&0.876&0.861&0.889&0.850&0.901\\
    2003.682&1.015&1.002&1.028&0.987&1.041&2013.890&0.888&0.873&0.900&0.861&0.913\\
    2003.756&0.995&0.981&1.009&0.968&1.022&2013.964&0.904&0.889&0.917&0.877&0.929\\
    2003.830&1.124&1.108&1.140&1.093&1.155&2014.038&0.901&0.887&0.914&0.874&0.927\\
    2003.904&1.168&1.153&1.181&1.138&1.196&2014.112&0.907&0.893&0.921&0.881&0.934\\
    2003.978&1.079&1.064&1.092&1.050&1.106&2014.186&0.927&0.912&0.939&0.899&0.952\\
    2004.052&1.046&1.031&1.058&1.017&1.072&2014.260&0.905&0.890&0.917&0.878&0.930\\
    2004.126&0.983&0.969&0.996&0.955&1.009&2014.334&0.872&0.859&0.885&0.846&0.897\\
    2004.200&0.935&0.922&0.947&0.908&0.960&2014.408&0.871&0.857&0.883&0.845&0.895\\
    2004.274&0.884&0.871&0.897&0.858&0.910&2014.482&0.880&0.866&0.893&0.853&0.905\\
    2004.348&0.856&0.843&0.868&0.830&0.880&2014.556&0.854&0.840&0.866&0.828&0.879\\
    2004.421&0.835&0.821&0.846&0.809&0.859&2014.630&0.798&0.784&0.811&0.772&0.823\\
    2004.495&0.809&0.796&0.820&0.783&0.833&2014.704&0.780&0.766&0.792&0.754&0.804\\
    2004.569&0.807&0.794&0.820&0.782&0.833&2014.778&0.757&0.744&0.769&0.733&0.781\\
    2004.643&0.787&0.774&0.799&0.761&0.811&2014.852&0.724&0.711&0.735&0.700&0.747\\
    2004.717&0.777&0.764&0.788&0.751&0.800&2014.926&0.739&0.725&0.751&0.715&0.763\\
    2004.791&0.743&0.730&0.755&0.718&0.767&2015.000&0.755&0.740&0.767&0.730&0.779\\
    2004.865&0.742&0.729&0.754&0.717&0.766&2015.074&0.757&0.744&0.769&0.732&0.781\\
    2004.939&0.725&0.712&0.736&0.700&0.747&2015.148&0.789&0.775&0.801&0.764&0.813\\
    2005.013&0.765&0.750&0.777&0.740&0.789&2015.221&0.835&0.820&0.847&0.809&0.860\\
    2005.087&0.780&0.768&0.792&0.756&0.804&2015.295&0.833&0.818&0.845&0.807&0.857\\
    2005.161&0.761&0.746&0.772&0.735&0.784&2015.369&0.792&0.778&0.803&0.766&0.815\\
    2005.235&0.750&0.738&0.762&0.725&0.773&2015.443&0.756&0.742&0.768&0.731&0.780\\
    2005.309&0.744&0.731&0.756&0.719&0.768&2015.517&0.726&0.713&0.739&0.703&0.751\\
    2005.383&0.741&0.727&0.752&0.716&0.763&2015.591&0.711&0.698&0.722&0.687&0.733\\
    2005.457&0.705&0.693&0.718&0.682&0.730&2015.665&0.701&0.687&0.712&0.676&0.723\\
    2005.531&0.730&0.715&0.740&0.703&0.752&2015.739&0.686&0.673&0.698&0.663&0.709\\
    2005.605&0.752&0.738&0.764&0.727&0.776&2015.813&0.682&0.669&0.694&0.658&0.705\\
    2005.679&0.797&0.783&0.810&0.771&0.823&2015.887&0.644&0.632&0.655&0.620&0.666\\
    2005.753&0.729&0.716&0.740&0.704&0.752&2015.961&0.608&0.596&0.620&0.586&0.630\\
    2005.827&0.683&0.672&0.694&0.660&0.706&2016.035&0.581&0.570&0.592&0.559&0.603\\
    2005.901&0.668&0.655&0.680&0.644&0.690&2016.109&0.561&0.550&0.572&0.540&0.582\\
    2005.975&0.652&0.640&0.663&0.629&0.674&2016.183&0.554&0.543&0.567&0.533&0.577\\
    2006.049&0.630&0.618&0.642&0.607&0.652&2016.257&0.553&0.542&0.564&0.531&0.575\\
    2006.123&0.610&0.597&0.621&0.586&0.631&2016.331&0.542&0.530&0.552&0.519&0.562\\
    2006.197&0.584&0.572&0.594&0.561&0.604&2016.405&0.528&0.516&0.537&0.506&0.547\\
    2006.271&0.573&0.561&0.584&0.550&0.594&2016.479&0.527&0.515&0.536&0.505&0.546\\
    2006.345&0.560&0.548&0.570&0.538&0.580&2016.553&0.520&0.508&0.530&0.498&0.540\\
    2006.419&0.550&0.538&0.560&0.528&0.570&2016.627&0.507&0.495&0.519&0.486&0.528\\
    2006.493&0.542&0.530&0.553&0.520&0.563&2016.701&0.494&0.482&0.503&0.472&0.513\\
    2006.567&0.540&0.528&0.549&0.517&0.559&2016.775&0.471&0.459&0.481&0.450&0.490\\
    2006.641&0.547&0.535&0.558&0.525&0.568&2016.849&0.454&0.442&0.463&0.433&0.472\\
    2006.715&0.528&0.517&0.538&0.506&0.548&2016.923&0.448&0.437&0.457&0.427&0.466\\
    2006.789&0.527&0.516&0.536&0.505&0.546&2016.997&0.436&0.425&0.449&0.416&0.456\\
    2006.863&0.517&0.505&0.527&0.495&0.536&2017.071&0.428&0.417&0.442&0.409&0.449\\
    2006.937&0.545&0.533&0.555&0.523&0.565&2017.145&0.420&0.409&0.432&0.400&0.440\\
    2007.011&0.514&0.503&0.525&0.493&0.534&2017.219&0.416&0.404&0.428&0.396&0.436\\
    2007.084&0.517&0.506&0.527&0.496&0.537&2017.293&0.413&0.402&0.424&0.393&0.432\\
    2007.158&0.504&0.493&0.514&0.483&0.523&2017.367&0.404&0.393&0.416&0.384&0.424\\
    2007.232&0.483&0.472&0.492&0.462&0.502&2017.441&0.395&0.384&0.405&0.375&0.413\\
    2007.306&0.474&0.462&0.483&0.453&0.492&2017.515&0.411&0.401&0.421&0.391&0.430\\
    2007.380&0.466&0.456&0.480&0.447&0.489&2017.589&0.435&0.424&0.445&0.415&0.454\\
    2007.454&0.463&0.452&0.474&0.443&0.483&2017.663&0.434&0.423&0.444&0.414&0.453\\
    2007.528&0.457&0.446&0.467&0.436&0.475&2017.737&0.459&0.447&0.469&0.439&0.478\\
    2007.602&0.455&0.445&0.467&0.436&0.476&2017.811&0.432&0.420&0.441&0.411&0.450\\
    2007.676&0.452&0.442&0.461&0.432&0.471&2017.884&0.404&0.392&0.415&0.384&0.423\\
    2007.750&0.447&0.436&0.458&0.427&0.466&2017.958&0.396&0.386&0.405&0.376&0.413\\
    \hline
    \label{tab:phi}
    \end{longtable}
\begin{table*}[ht!]
    \centering
    \caption{\upshape LIS fluxes of Helium nuclei based on AMS-02 
    measurements.}
    \begin{tabular}{ccc||ccc}
    \toprule
    
    $E_k$ (GeV/n) & Flux (m$^{-2}$s$^{-1}$sr$^{-1}$(GeV/n)$^{-1}$) & $\sigma$ &
    $E_k$ (GeV/n) & Flux (m$^{-2}$s$^{-1}$sr$^{-1}$(GeV/n)$^{-1}$) & $\sigma$\\
    \cline{1-6}

    7.730e-01&3.223e+02&7.504e+00&1.960e+01&2.459e-01&3.341e-03\\
    8.633e-01&2.650e+02&4.424e+00&2.115e+01&2.008e-01&2.761e-03\\
    9.657e-01&2.153e+02&3.208e+00&2.279e+01&1.649e-01&2.281e-03\\
    1.082e+00&1.770e+02&2.510e+00&2.457e+01&1.342e-01&1.862e-03\\
    1.213e+00&1.455e+02&2.086e+00&2.646e+01&1.103e-01&1.530e-03\\
    1.358e+00&1.190e+02&1.700e+00&2.849e+01&9.052e-02&1.260e-03\\
    1.521e+00&9.702e+01&1.427e+00&3.066e+01&7.423e-02&1.053e-03\\
    1.699e+00&7.886e+01&1.164e+00&3.301e+01&6.092e-02&8.661e-04\\
    1.897e+00&6.378e+01&9.578e-01&3.553e+01&5.007e-02&7.005e-04\\
    2.116e+00&5.137e+01&7.484e-01&3.823e+01&4.110e-02&5.758e-04\\
    2.357e+00&4.121e+01&6.092e-01&4.112e+01&3.365e-02&4.746e-04\\
    2.619e+00&3.305e+01&4.827e-01&4.425e+01&2.765e-02&3.906e-04\\
    2.904e+00&2.648e+01&3.663e-01&4.762e+01&2.281e-02&3.302e-04\\
    3.216e+00&2.120e+01&2.971e-01&5.136e+01&1.857e-02&2.682e-04\\
    3.553e+00&1.701e+01&2.357e-01&5.536e+01&1.512e-02&2.215e-04\\
    3.919e+00&1.365e+01&1.871e-01&5.961e+01&1.232e-02&1.801e-04\\
    4.316e+00&1.098e+01&1.480e-01&6.435e+01&1.002e-02&1.490e-04\\
    4.743e+00&8.823e+00&1.174e-01&6.983e+01&7.997e-03&1.200e-04\\
    5.211e+00&7.078e+00&9.482e-02&7.608e+01&6.346e-03&9.542e-05\\
    5.705e+00&5.720e+00&7.541e-02&8.306e+01&5.017e-03&7.677e-05\\
    6.224e+00&4.634e+00&6.105e-02&9.105e+01&3.886e-03&6.061e-05\\
    6.794e+00&3.746e+00&4.913e-02&1.000e+02&3.028e-03&4.786e-05\\
    7.414e+00&3.027e+00&3.983e-02&1.103e+02&2.306e-03&3.759e-05\\
    8.084e+00&2.439e+00&3.291e-02&1.222e+02&1.754e-03&2.997e-05\\
    8.805e+00&1.971e+00&2.634e-02&1.367e+02&1.297e-03&2.283e-05\\
    9.576e+00&1.589e+00&2.119e-02&1.543e+02&9.386e-04&1.741e-05\\
    1.040e+01&1.288e+00&1.719e-02&1.762e+02&6.623e-04&1.302e-05\\
    1.129e+01&1.044e+00&1.410e-02&2.038e+02&4.456e-04&9.398e-06\\
    1.226e+01&8.429e-01&1.127e-02&2.400e+02&2.898e-04&6.667e-06\\
    1.329e+01&6.855e-01&9.302e-03&2.899e+02&1.782e-04&4.543e-06\\
    1.438e+01&5.572e-01&7.587e-03&3.629e+02&9.730e-05&2.863e-06\\
    1.556e+01&4.535e-01&6.082e-03&4.813e+02&4.742e-05&1.710e-06\\
    1.680e+01&3.701e-01&5.021e-03&7.125e+02&1.710e-05&8.520e-07\\
    1.815e+01&3.021e-01&4.177e-03&1.161e+03&4.929e-06&3.954e-07\\
    \hline
    \end{tabular}

    \label{tab:He}
\end{table*}

\begin{table*}[ht!]
    \centering
    \caption{\upshape LIS fluxes of Lithium nuclei based on AMS-02
    measurements.}
    \begin{tabular}{ccc||ccc}
    \toprule

    $E_k$ (GeV/n) & Flux (m$^{-2}$s$^{-1}$sr$^{-1}$(GeV/n)$^{-1}$) & $\sigma$ &
    $E_k$ (GeV/n) & Flux (m$^{-2}$s$^{-1}$sr$^{-1}$(GeV/n)$^{-1}$) & $\sigma$\\
    
    \cline{1-6}
    7.076e-01&1.543e+00&5.057e-02&1.804e+01&1.165e-03&3.319e-05\\
    7.881e-01&1.408e+00&4.236e-02&1.947e+01&9.221e-04&2.667e-05\\
    8.796e-01&1.254e+00&3.602e-02&2.099e+01&7.464e-04&2.175e-05\\
    9.841e-01&1.093e+00&3.067e-02&2.262e+01&5.738e-04&1.707e-05\\
    1.102e+00&9.361e-01&2.593e-02&2.438e+01&4.633e-04&1.403e-05\\
    1.234e+00&7.840e-01&2.168e-02&2.624e+01&3.685e-04&1.133e-05\\
    1.381e+00&6.493e-01&1.770e-02&2.825e+01&2.998e-04&9.324e-06\\
    1.543e+00&5.378e-01&1.461e-02&3.041e+01&2.345e-04&7.440e-06\\
    1.723e+00&4.405e-01&1.189e-02&3.274e+01&1.887e-04&6.173e-06\\
    1.923e+00&3.595e-01&9.775e-03&3.523e+01&1.472e-04&4.964e-06\\
    2.143e+00&2.910e-01&7.829e-03&3.791e+01&1.175e-04&4.095e-06\\
    2.383e+00&2.340e-01&6.304e-03&4.079e+01&9.648e-05&3.434e-06\\
    2.645e+00&1.883e-01&5.064e-03&4.390e+01&7.270e-05&2.712e-06\\
    2.931e+00&1.501e-01&4.046e-03&4.735e+01&5.888e-05&2.240e-06\\
    3.241e+00&1.194e-01&3.219e-03&5.105e+01&4.454e-05&1.815e-06\\
    3.577e+00&9.507e-02&2.546e-03&5.497e+01&3.511e-05&1.488e-06\\
    3.942e+00&7.558e-02&2.032e-03&5.934e+01&2.827e-05&1.240e-06\\
    4.336e+00&5.976e-02&1.616e-03&6.441e+01&2.341e-05&1.038e-06\\
    4.766e+00&4.687e-02&1.266e-03&7.017e+01&1.819e-05&8.424e-07\\
    5.222e+00&3.763e-02&1.026e-03&7.662e+01&1.398e-05&6.756e-07\\
    5.700e+00&2.973e-02&7.948e-04&8.399e+01&1.019e-05&5.253e-07\\
    6.225e+00&2.382e-02&6.456e-04&9.228e+01&7.549e-06&4.165e-07\\
    6.796e+00&1.882e-02&5.120e-04&1.017e+02&5.542e-06&3.253e-07\\
    7.414e+00&1.476e-02&4.020e-04&1.128e+02&4.500e-06&2.701e-07\\
    8.079e+00&1.172e-02&3.204e-04&1.261e+02&2.783e-06&1.867e-07\\
    8.790e+00&9.259e-03&2.541e-04&1.424e+02&2.029e-06&1.431e-07\\
    9.548e+00&7.381e-03&2.029e-04&1.626e+02&1.558e-06&1.130e-07\\
    1.037e+01&5.856e-03&1.614e-04&1.881e+02&1.081e-06&8.370e-08\\
    1.127e+01&4.596e-03&1.254e-04&2.215e+02&4.787e-07&4.590e-08\\
    1.221e+01&3.663e-03&1.015e-04&2.711e+02&3.416e-07&3.223e-08\\
    1.323e+01&2.940e-03&8.011e-05&3.511e+02&1.555e-07&1.739e-08\\
    1.431e+01&2.308e-03&6.389e-05&4.930e+02&5.949e-08&8.211e-09\\
    1.546e+01&1.831e-03&5.135e-05&9.553e+02&1.127e-08&3.001e-09\\
    1.670e+01&1.449e-03&4.083e-05&&&\\
    
    \hline
    \end{tabular}

    \label{tab:Li}
\end{table*}
\begin{table*}[ht!]
    \centering
    \caption{\upshape LIS fluxes of Beryllium nuclei based on AMS-02
    measurements.}
    \begin{tabular}{ccc||ccc}
    \toprule
    
    $E_k$ (GeV/n) & Flux (m$^{-2}$s$^{-1}$sr$^{-1}$(GeV/n)$^{-1}$) & $\sigma$ &
    $E_k$ (GeV/n) & Flux (m$^{-2}$s$^{-1}$sr$^{-1}$(GeV/n)$^{-1}$) & $\sigma$\\
    \cline{1-6}

    7.918e-01&7.483e-01&2.587e-02&1.962e+01&5.403e-04&1.706e-05\\
    8.822e-01&6.571e-01&2.170e-02&2.116e+01&4.205e-04&1.360e-05\\
    9.845e-01&5.549e-01&1.764e-02&2.281e+01&3.443e-04&1.136e-05\\
    1.101e+00&4.788e-01&1.475e-02&2.458e+01&2.717e-04&9.141e-06\\
    1.232e+00&4.124e-01&1.246e-02&2.648e+01&2.170e-04&7.576e-06\\
    1.377e+00&3.415e-01&1.022e-02&2.851e+01&1.743e-04&6.260e-06\\
    1.539e+00&2.843e-01&8.307e-03&3.068e+01&1.417e-04&5.210e-06\\
    1.718e+00&2.326e-01&6.746e-03&3.302e+01&1.077e-04&4.141e-06\\
    1.916e+00&1.907e-01&5.471e-03&3.555e+01&9.018e-05&3.546e-06\\
    2.135e+00&1.563e-01&4.445e-03&3.825e+01&6.820e-05&2.840e-06\\
    2.376e+00&1.253e-01&3.547e-03&4.114e+01&5.721e-05&2.446e-06\\
    2.638e+00&1.011e-01&2.870e-03&4.426e+01&4.805e-05&2.097e-06\\
    2.923e+00&8.098e-02&2.299e-03&4.764e+01&3.705e-05&1.709e-06\\
    3.235e+00&6.439e-02&1.812e-03&5.138e+01&2.842e-05&1.379e-06\\
    3.572e+00&5.051e-02&1.425e-03&5.538e+01&2.288e-05&1.180e-06\\
    3.938e+00&4.063e-02&1.151e-03&5.962e+01&1.792e-05&9.681e-07\\
    4.335e+00&3.239e-02&9.184e-04&6.437e+01&1.275e-05&7.511e-07\\
    4.762e+00&2.562e-02&7.132e-04&6.985e+01&1.133e-05&6.505e-07\\
    5.230e+00&1.998e-02&5.666e-04&7.610e+01&7.819e-06&4.999e-07\\
    5.724e+00&1.626e-02&4.615e-04&8.308e+01&6.671e-06&4.296e-07\\
    6.243e+00&1.290e-02&3.681e-04&9.107e+01&4.337e-06&3.145e-07\\
    6.813e+00&1.024e-02&2.943e-04&1.001e+02&3.212e-06&2.516e-07\\
    7.433e+00&8.156e-03&2.350e-04&1.103e+02&2.904e-06&2.227e-07\\
    8.103e+00&6.598e-03&1.908e-04&1.222e+02&2.042e-06&1.702e-07\\
    8.823e+00&5.157e-03&1.496e-04&1.367e+02&1.500e-06&1.310e-07\\
    9.595e+00&4.152e-03&1.219e-04&1.544e+02&9.605e-07&9.363e-08\\
    1.042e+01&3.333e-03&9.685e-05&1.762e+02&6.176e-07&6.654e-08\\
    1.131e+01&2.615e-03&7.612e-05&2.038e+02&4.804e-07&5.279e-08\\
    1.228e+01&2.044e-03&6.058e-05&2.400e+02&2.469e-07&3.220e-08\\
    1.331e+01&1.638e-03&4.891e-05&2.937e+02&1.552e-07&2.099e-08\\
    1.440e+01&1.318e-03&3.948e-05&3.805e+02&7.219e-08&1.165e-08\\
    1.558e+01&1.069e-03&3.249e-05&5.342e+02&2.503e-08&5.280e-09\\
    1.682e+01&8.429e-04&2.588e-05&1.035e+03&2.802e-09&1.476e-09\\
    1.817e+01&6.771e-04&2.121e-05&&&\\     
    \hline
    \end{tabular}

    \label{tab:Be}
\end{table*}

\begin{table*}[ht!]
    \centering
    \caption{\upshape LIS fluxes of Boron nuclei based on ACE and AMS-02
    measurements.}
    \begin{tabular}{ccc||ccc}
    \toprule
    
    $E_k$ (GeV/n) & Flux (m$^{-2}$s$^{-1}$sr$^{-1}$(GeV/n)$^{-1}$) & $\sigma$ &
    $E_k$ (GeV/n) & Flux (m$^{-2}$s$^{-1}$sr$^{-1}$(GeV/n)$^{-1}$) & $\sigma$\\
    \cline{1-6}
    3.848e-01&4.507e+00&1.427e-01&1.340e+01&3.960e-03&9.973e-05\\
    4.049e-01&4.084e+00&1.289e-01&1.450e+01&3.144e-03&8.148e-05\\
    4.272e-01&4.041e+00&1.334e-01&1.566e+01&2.543e-03&6.497e-05\\
    4.463e-01&3.922e+00&1.383e-01&1.692e+01&2.008e-03&5.283e-05\\
    4.634e-01&3.799e+00&1.456e-01&1.828e+01&1.566e-03&4.208e-05\\
    4.792e-01&3.655e+00&1.510e-01&1.972e+01&1.264e-03&3.428e-05\\
    4.938e-01&3.579e+00&1.635e-01&2.126e+01&1.030e-03&2.836e-05\\
    \cline{1-3}
    7.219e-01&2.585e+00&1.488e-01&2.292e+01&7.982e-04&2.261e-05\\
    8.038e-01&2.322e+00&1.188e-01&2.469e+01&6.452e-04&1.849e-05\\
    8.969e-01&2.050e+00&9.443e-02&2.658e+01&5.136e-04&1.501e-05\\
    1.003e+00&1.726e+00&7.201e-02&2.861e+01&4.065e-04&1.224e-05\\
    1.123e+00&1.433e+00&5.459e-02&3.081e+01&3.232e-04&9.874e-06\\
    1.257e+00&1.186e+00&4.194e-02&3.316e+01&2.579e-04&8.310e-06\\
    1.406e+00&9.754e-01&3.189e-02&3.568e+01&2.104e-04&6.844e-06\\
    1.571e+00&7.956e-01&2.444e-02&3.839e+01&1.676e-04&5.609e-06\\
    1.754e+00&6.365e-01&1.881e-02&4.131e+01&1.369e-04&4.697e-06\\
    1.957e+00&5.166e-01&1.466e-02&4.446e+01&1.054e-04&3.779e-06\\
    2.180e+00&4.114e-01&1.118e-02&4.796e+01&8.215e-05&3.067e-06\\
    2.423e+00&3.276e-01&8.538e-03&5.170e+01&6.425e-05&2.539e-06\\
    2.688e+00&2.605e-01&6.756e-03&5.566e+01&5.476e-05&2.182e-06\\
    2.978e+00&2.057e-01&5.240e-03&6.010e+01&4.073e-05&1.716e-06\\
    3.292e+00&1.629e-01&4.096e-03&6.522e+01&3.199e-05&1.377e-06\\
    3.633e+00&1.293e-01&3.212e-03&7.106e+01&2.337e-05&1.084e-06\\
    4.003e+00&1.022e-01&2.523e-03&7.759e+01&1.885e-05&8.957e-07\\
    4.401e+00&8.142e-02&2.009e-03&8.505e+01&1.511e-05&7.397e-07\\
    4.837e+00&6.378e-02&1.562e-03&9.345e+01&1.094e-05&5.770e-07\\
    5.299e+00&5.092e-02&1.252e-03&1.030e+02&8.194e-06&4.581e-07\\
    5.783e+00&4.041e-02&9.977e-04&1.142e+02&6.151e-06&3.604e-07\\
    6.314e+00&3.216e-02&7.967e-04&1.277e+02&3.990e-06&2.583e-07\\
    6.893e+00&2.565e-02&6.233e-04&1.442e+02&3.059e-06&2.048e-07\\
    7.519e+00&2.005e-02&4.928e-04&1.646e+02&1.905e-06&1.415e-07\\
    8.192e+00&1.615e-02&3.979e-04&1.904e+02&1.291e-06&1.030e-07\\
    8.912e+00&1.274e-02&3.158e-04&2.242e+02&8.297e-07&7.203e-08\\
    9.679e+00&1.009e-02&2.511e-04&2.745e+02&4.962e-07&4.630e-08\\
    1.052e+01&7.995e-03&1.984e-04&3.555e+02&2.161e-07&2.441e-08\\
    1.142e+01&6.366e-03&1.605e-04&4.992e+02&9.469e-08&1.307e-08\\
    1.238e+01&5.121e-03&1.302e-04&9.673e+02&1.129e-08&3.329e-09\\

    \hline
    \end{tabular}

    \label{tab:B}
\end{table*}

\begin{table*}[ht!]
    \centering
    \caption{\upshape LIS fluxes of Carbon nuclei based on ACE and AMS-02
    measurements.}
    \begin{tabular}{ccc||ccc}
    \toprule
    
    $E_k$ (GeV/n) & Flux (m$^{-2}$s$^{-1}$sr$^{-1}$(GeV/n)$^{-1}$) & $\sigma$ &
    $E_k$ (GeV/n) & Flux (m$^{-2}$s$^{-1}$sr$^{-1}$(GeV/n)$^{-1}$) & $\sigma$\\
    \cline{1-6}
    4.163e-01&1.704e+01&5.323e-01&1.558e+01&1.460e-02&3.482e-04\\
    4.395e-01&1.624e+01&5.143e-01&1.682e+01&1.197e-02&2.863e-04\\
    4.653e-01&1.578e+01&5.246e-01&1.817e+01&9.798e-03&2.351e-04\\
    4.873e-01&1.494e+01&5.325e-01&1.962e+01&8.051e-03&1.949e-04\\
    5.071e-01&1.445e+01&5.505e-01&2.117e+01&6.522e-03&1.594e-04\\
    5.254e-01&1.399e+01&5.869e-01&2.281e+01&5.342e-03&1.298e-04\\
    5.425e-01&1.324e+01&6.163e-01&2.459e+01&4.458e-03&1.090e-04\\
    \cline{1-3}
    7.927e-01&8.002e+00&2.863e-01&2.648e+01&3.624e-03&9.021e-05\\
    8.830e-01&7.069e+00&2.121e-01&2.851e+01&2.988e-03&7.315e-05\\
    9.854e-01&6.099e+00&1.716e-01&3.068e+01&2.445e-03&6.104e-05\\
    1.102e+00&5.071e+00&1.353e-01&3.303e+01&1.986e-03&4.994e-05\\
    1.233e+00&4.117e+00&1.053e-01&3.555e+01&1.632e-03&4.145e-05\\
    1.378e+00&3.349e+00&8.414e-02&3.825e+01&1.350e-03&3.453e-05\\
    1.540e+00&2.717e+00&6.679e-02&4.114e+01&1.097e-03&2.814e-05\\
    1.719e+00&2.230e+00&5.390e-02&4.427e+01&9.098e-04&2.353e-05\\
    1.917e+00&1.824e+00&4.368e-02&4.764e+01&7.365e-04&1.938e-05\\
    2.136e+00&1.493e+00&3.536e-02&5.138e+01&6.059e-04&1.596e-05\\
    2.377e+00&1.212e+00&2.834e-02&5.538e+01&4.885e-04&1.313e-05\\
    2.638e+00&9.846e-01&2.296e-02&5.963e+01&4.057e-04&1.101e-05\\
    2.924e+00&7.980e-01&1.848e-02&6.437e+01&3.239e-04&8.898e-06\\
    3.236e+00&6.424e-01&1.477e-02&6.985e+01&2.600e-04&7.165e-06\\
    3.573e+00&5.197e-01&1.200e-02&7.610e+01&2.024e-04&5.698e-06\\
    3.939e+00&4.179e-01&9.615e-03&8.308e+01&1.606e-04&4.505e-06\\
    4.336e+00&3.358e-01&7.790e-03&9.107e+01&1.262e-04&3.606e-06\\
    4.763e+00&2.701e-01&6.283e-03&1.001e+02&9.592e-05&2.801e-06\\
    5.231e+00&2.165e-01&5.008e-03&1.103e+02&7.545e-05&2.251e-06\\
    5.725e+00&1.753e-01&4.043e-03&1.222e+02&5.765e-05&1.755e-06\\
    6.244e+00&1.425e-01&3.306e-03&1.367e+02&4.376e-05&1.377e-06\\
    6.814e+00&1.153e-01&2.674e-03&1.544e+02&3.086e-05&9.993e-07\\
    7.433e+00&9.378e-02&2.166e-03&1.762e+02&2.227e-05&7.646e-07\\
    8.104e+00&7.577e-02&1.754e-03&2.038e+02&1.434e-05&5.221e-07\\
    8.824e+00&6.160e-02&1.437e-03&2.400e+02&9.800e-06&3.836e-07\\
    9.595e+00&5.038e-02&1.169e-03&2.899e+02&5.916e-06&2.602e-07\\
    1.042e+01&4.093e-02&9.685e-04&3.629e+02&3.252e-06&1.685e-07\\
    1.131e+01&3.325e-02&7.707e-04&4.813e+02&1.514e-06&9.593e-08\\
    1.228e+01&2.693e-02&6.408e-04&7.125e+02&5.571e-07&5.034e-08\\
    1.331e+01&2.193e-02&5.098e-04&1.161e+03&1.733e-07&2.217e-08\\
    1.440e+01&1.786e-02&4.231e-04&0.000e+00&0.000e+00&0.000e+00\\

    \hline
    \end{tabular}

    \label{tab:C}
\end{table*}

\begin{table*}[ht!]
    \centering
    \caption{\upshape LIS fluxes of Oxygen nuclei based on ACE and AMS-02
    measurements.}
    \begin{tabular}{ccc||ccc}
    \toprule
    
    $E_k$ (GeV/n) & Flux (m$^{-2}$s$^{-1}$sr$^{-1}$(GeV/n)$^{-1}$) & $\sigma$ &
    $E_k$ (GeV/n) & Flux (m$^{-2}$s$^{-1}$sr$^{-1}$(GeV/n)$^{-1}$) & $\sigma$\\
    \cline{1-6}
    4.284e-01&1.646e+01&5.071e-01&1.558e+01&1.540e-02&4.317e-04\\
    4.558e-01&1.570e+01&4.921e-01&1.682e+01&1.265e-02&3.571e-04\\
    4.864e-01&1.496e+01&4.820e-01&1.817e+01&1.037e-02&2.933e-04\\
    5.127e-01&1.397e+01&4.651e-01&1.962e+01&8.451e-03&2.406e-04\\
    5.364e-01&1.345e+01&4.688e-01&2.117e+01&6.970e-03&1.985e-04\\
    5.583e-01&1.281e+01&4.623e-01&2.281e+01&5.738e-03&1.648e-04\\
    5.788e-01&1.189e+01&4.432e-01&2.459e+01&4.760e-03&1.377e-04\\
    \cline{1-3}
    8.830e-01&6.785e+00&2.359e-01&2.648e+01&3.919e-03&1.148e-04\\
    9.854e-01&5.674e+00&1.838e-01&2.851e+01&3.191e-03&9.367e-05\\
    1.102e+00&4.713e+00&1.465e-01&3.068e+01&2.639e-03&7.704e-05\\
    1.233e+00&3.897e+00&1.157e-01&3.303e+01&2.173e-03&6.449e-05\\
    1.378e+00&3.192e+00&9.235e-02&3.555e+01&1.756e-03&5.261e-05\\
    1.540e+00&2.611e+00&7.364e-02&3.825e+01&1.458e-03&4.412e-05\\
    1.719e+00&2.128e+00&5.923e-02&4.114e+01&1.207e-03&3.685e-05\\
    1.917e+00&1.738e+00&4.791e-02&4.427e+01&9.796e-04&3.021e-05\\
    2.136e+00&1.414e+00&3.861e-02&4.764e+01&8.000e-04&2.481e-05\\
    2.377e+00&1.148e+00&3.126e-02&5.138e+01&6.570e-04&2.061e-05\\
    2.638e+00&9.314e-01&2.526e-02&5.538e+01&5.420e-04&1.736e-05\\
    2.924e+00&7.536e-01&2.047e-02&5.963e+01&4.415e-04&1.421e-05\\
    3.236e+00&6.080e-01&1.648e-02&6.437e+01&3.601e-04&1.170e-05\\
    3.573e+00&4.937e-01&1.345e-02&6.985e+01&2.927e-04&9.582e-06\\
    3.939e+00&4.000e-01&1.080e-02&7.610e+01&2.265e-04&7.536e-06\\
    4.336e+00&3.238e-01&8.737e-03&8.308e+01&1.796e-04&6.107e-06\\
    4.763e+00&2.616e-01&7.202e-03&9.107e+01&1.418e-04&4.899e-06\\
    5.231e+00&2.119e-01&5.758e-03&1.001e+02&1.124e-04&3.957e-06\\
    5.725e+00&1.731e-01&4.718e-03&1.103e+02&8.431e-05&3.039e-06\\
    6.244e+00&1.415e-01&3.863e-03&1.222e+02&6.376e-05&2.351e-06\\
    6.814e+00&1.146e-01&3.136e-03&1.367e+02&4.764e-05&1.808e-06\\
    7.433e+00&9.411e-02&2.582e-03&1.544e+02&3.423e-05&1.348e-06\\
    8.104e+00&7.662e-02&2.102e-03&1.762e+02&2.456e-05&9.971e-07\\
    8.824e+00&6.264e-02&1.719e-03&2.038e+02&1.650e-05&7.180e-07\\
    9.595e+00&5.118e-02&1.406e-03&2.400e+02&1.054e-05&4.925e-07\\
    1.042e+01&4.178e-02&1.162e-03&2.899e+02&6.259e-06&3.228e-07\\
    1.131e+01&3.429e-02&9.420e-04&3.629e+02&3.507e-06&2.071e-07\\
    1.228e+01&2.791e-02&7.667e-04&4.813e+02&1.738e-06&1.217e-07\\
    1.331e+01&2.277e-02&6.354e-04&7.125e+02&6.594e-07&6.318e-08\\
    1.440e+01&1.873e-02&5.239e-04&1.161e+03&1.933e-07&2.588e-08\\

    \hline
    \end{tabular}

    \label{tab:O}
\end{table*}

\begin{figure}[ht!]
\centering
\includegraphics[scale=0.3]{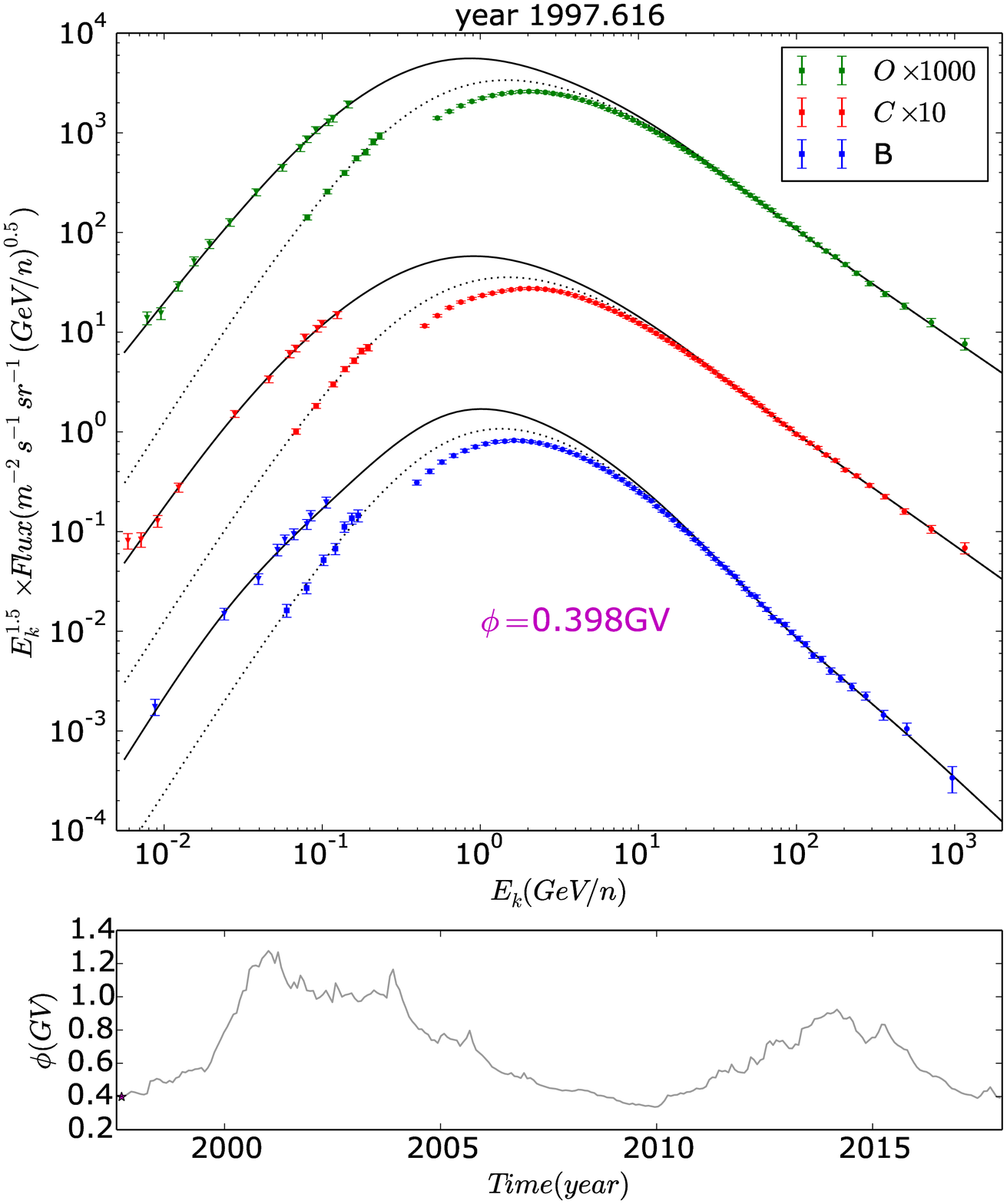}
\includegraphics[scale=0.3]{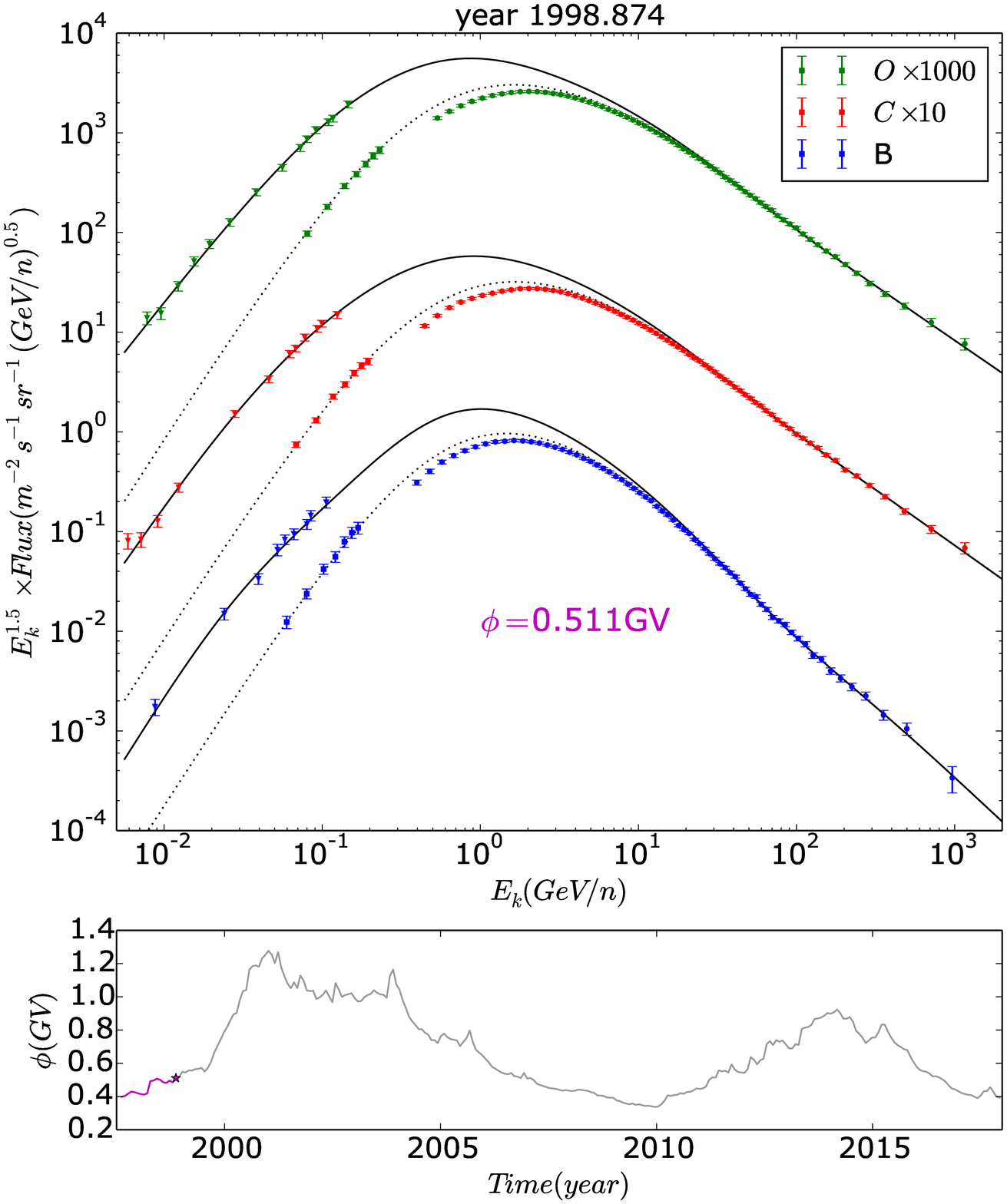}
\includegraphics[scale=0.3]{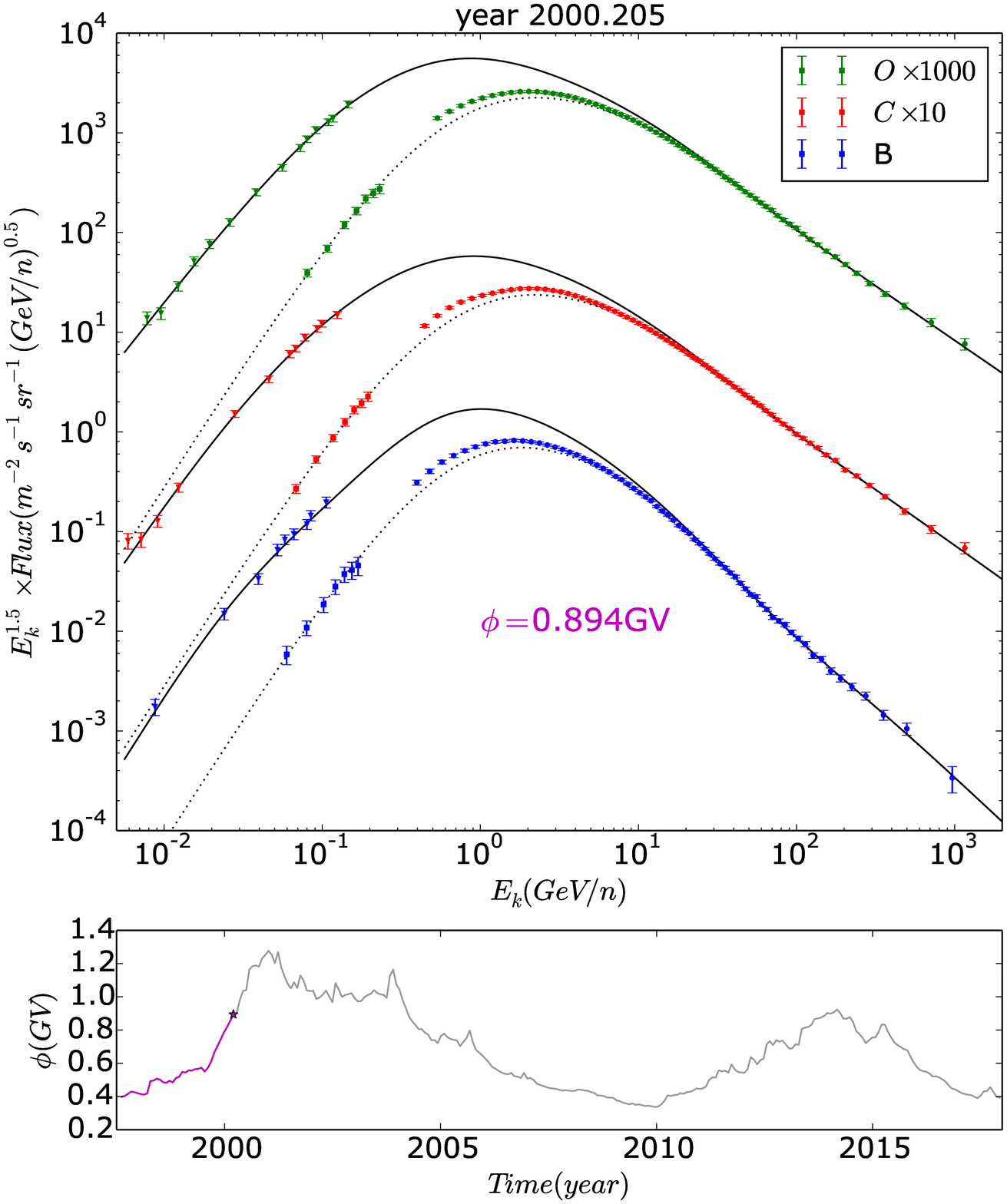}
\includegraphics[scale=0.3]{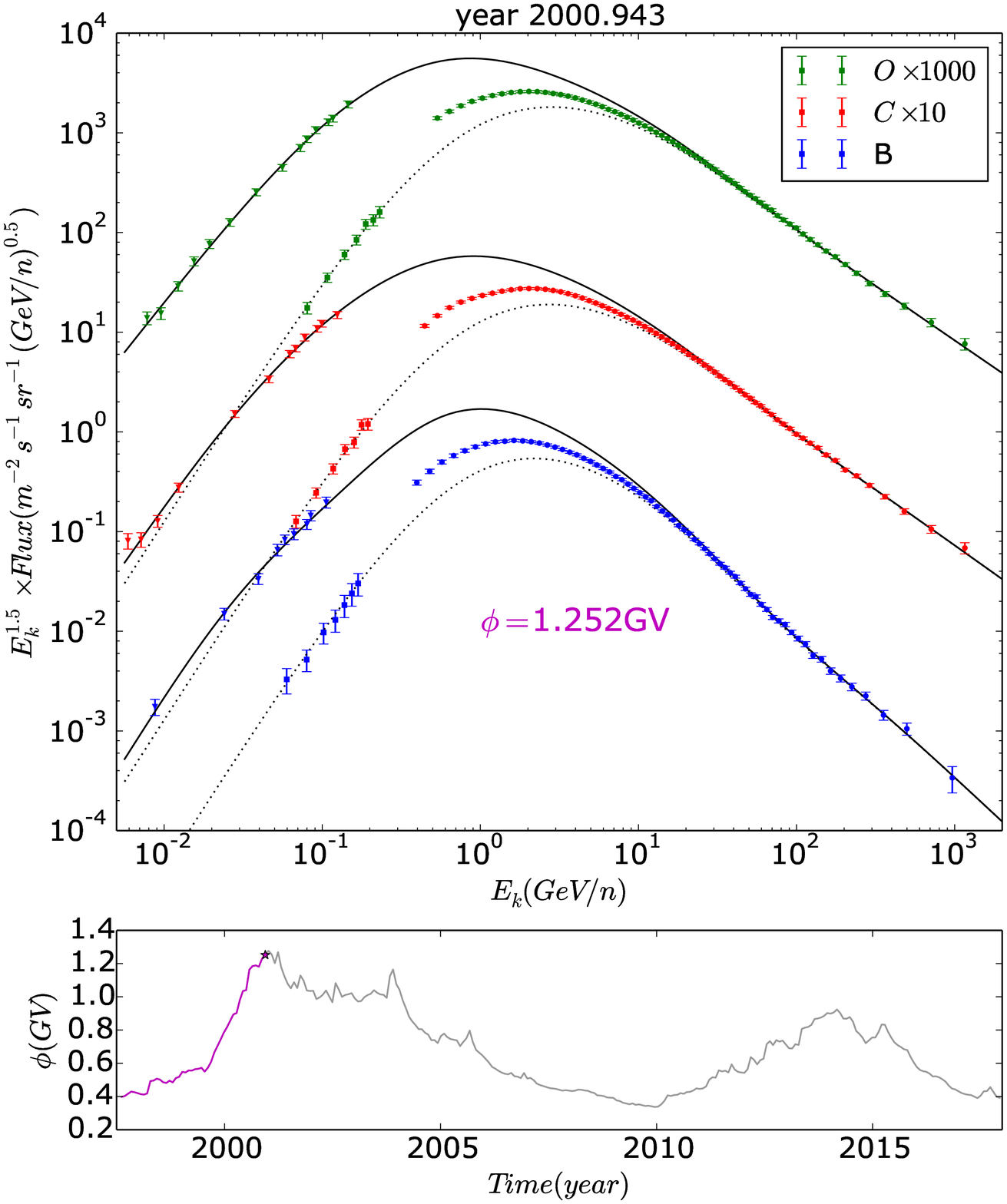}
\caption{Fluxes of Boron, Carbon, and Oxygen nuclei. The Voyager-1 data
represent the LIS fluxes, and the ACE and AMS-02 data are the TOA fluxes.
Different panels are for different time, for which the ACE data are
different. The solid lines are the best-fit LIS results, and the dotted
lines are the model fluxes to fit the ACE data at this particular time,
with the modulation potential labelled in the plot.
The fluxes are multiplied by $E^{1.5}$.}
\label{Fig.1}
\end{figure}

\begin{figure}[ht!]
\centering
\includegraphics[scale=0.3]{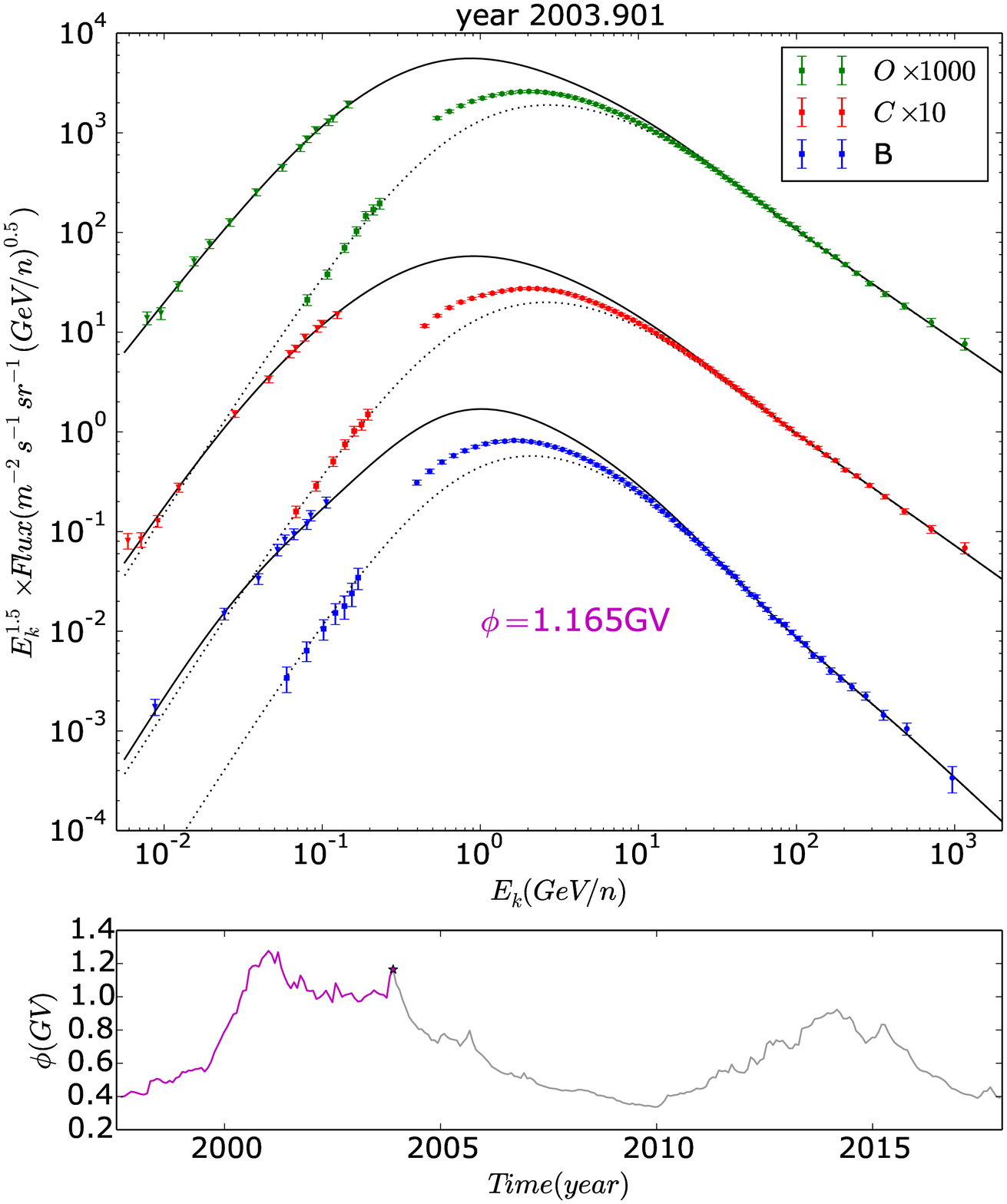}
\includegraphics[scale=0.3]{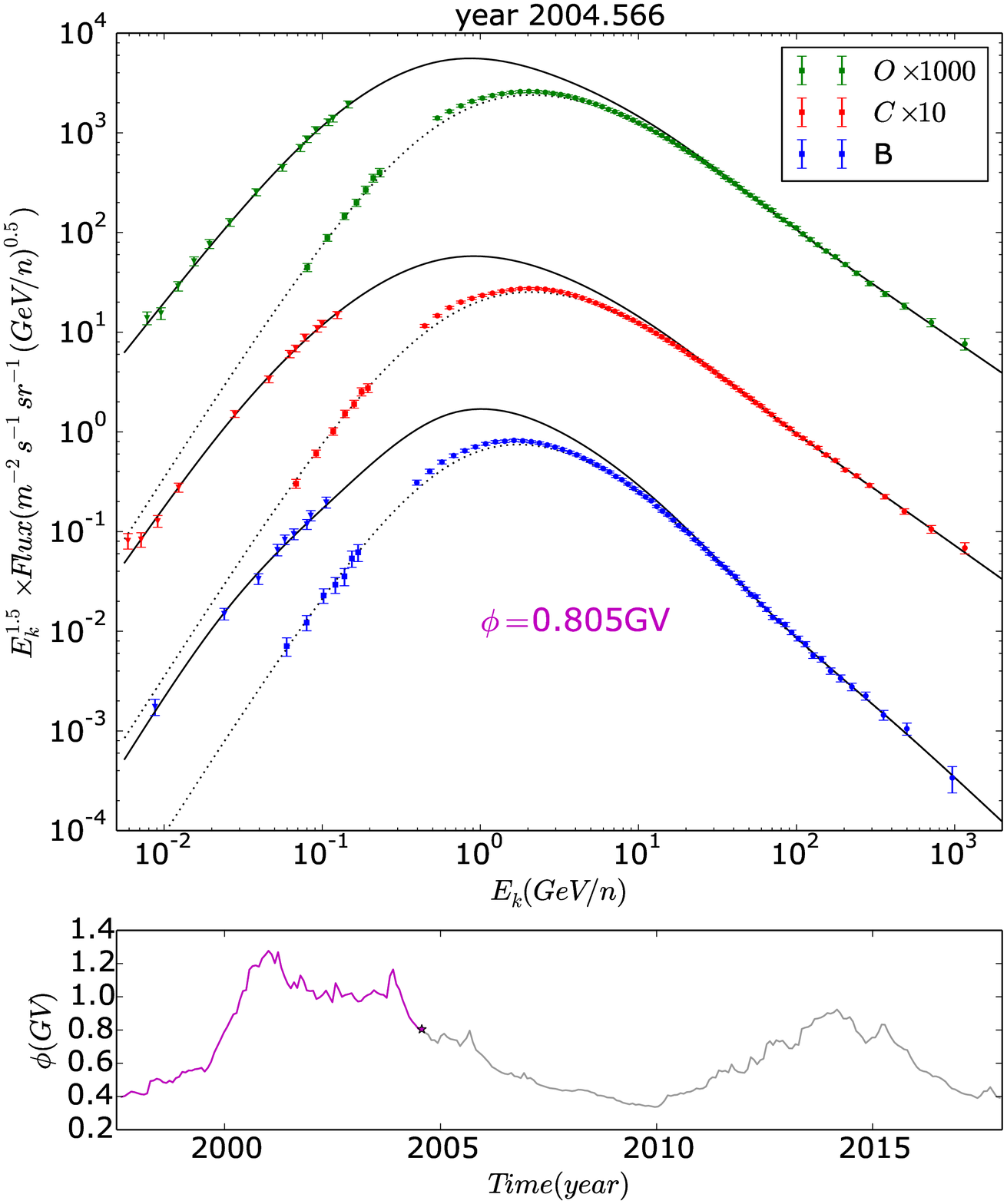}
\includegraphics[scale=0.3]{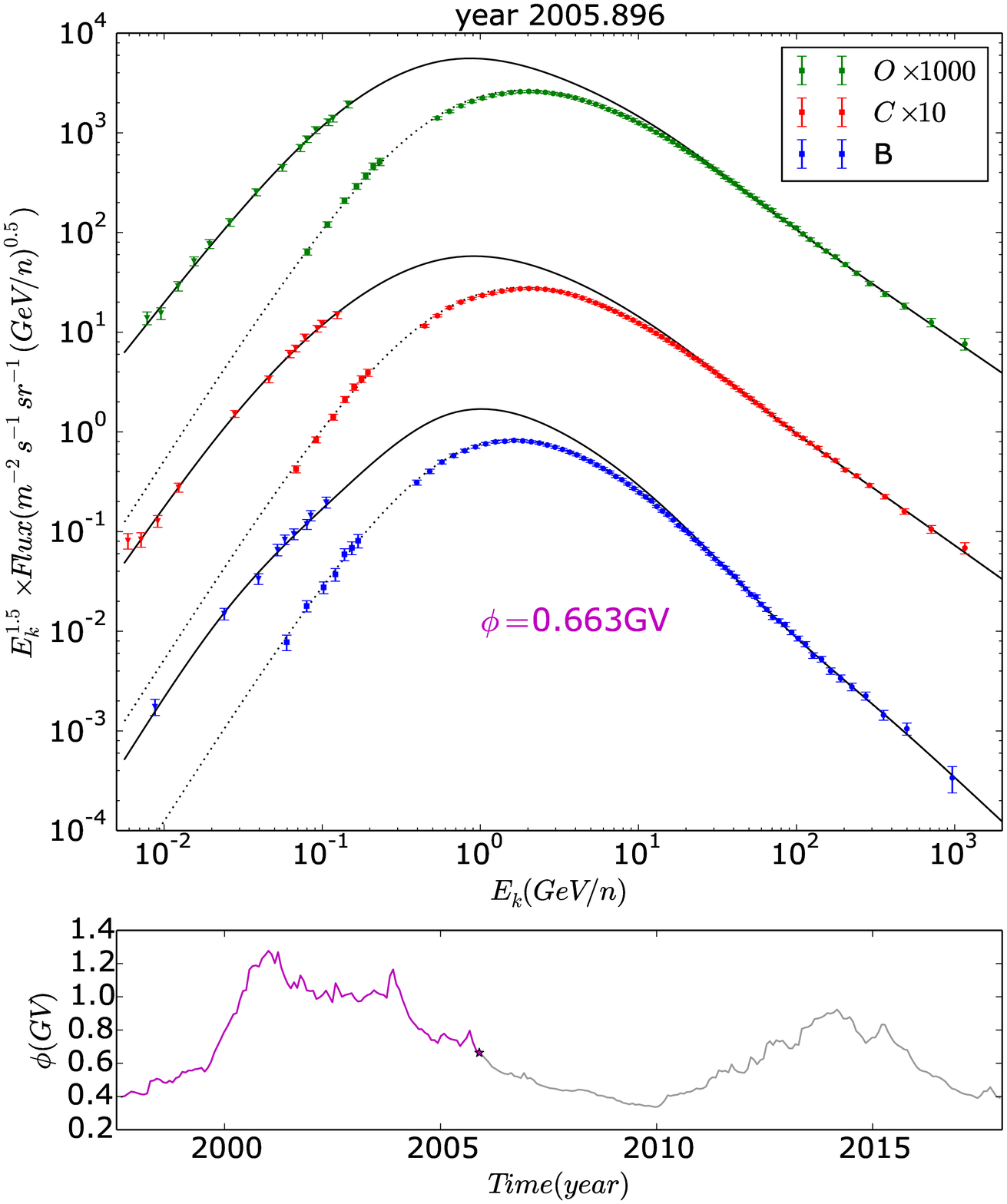}
\includegraphics[scale=0.3]{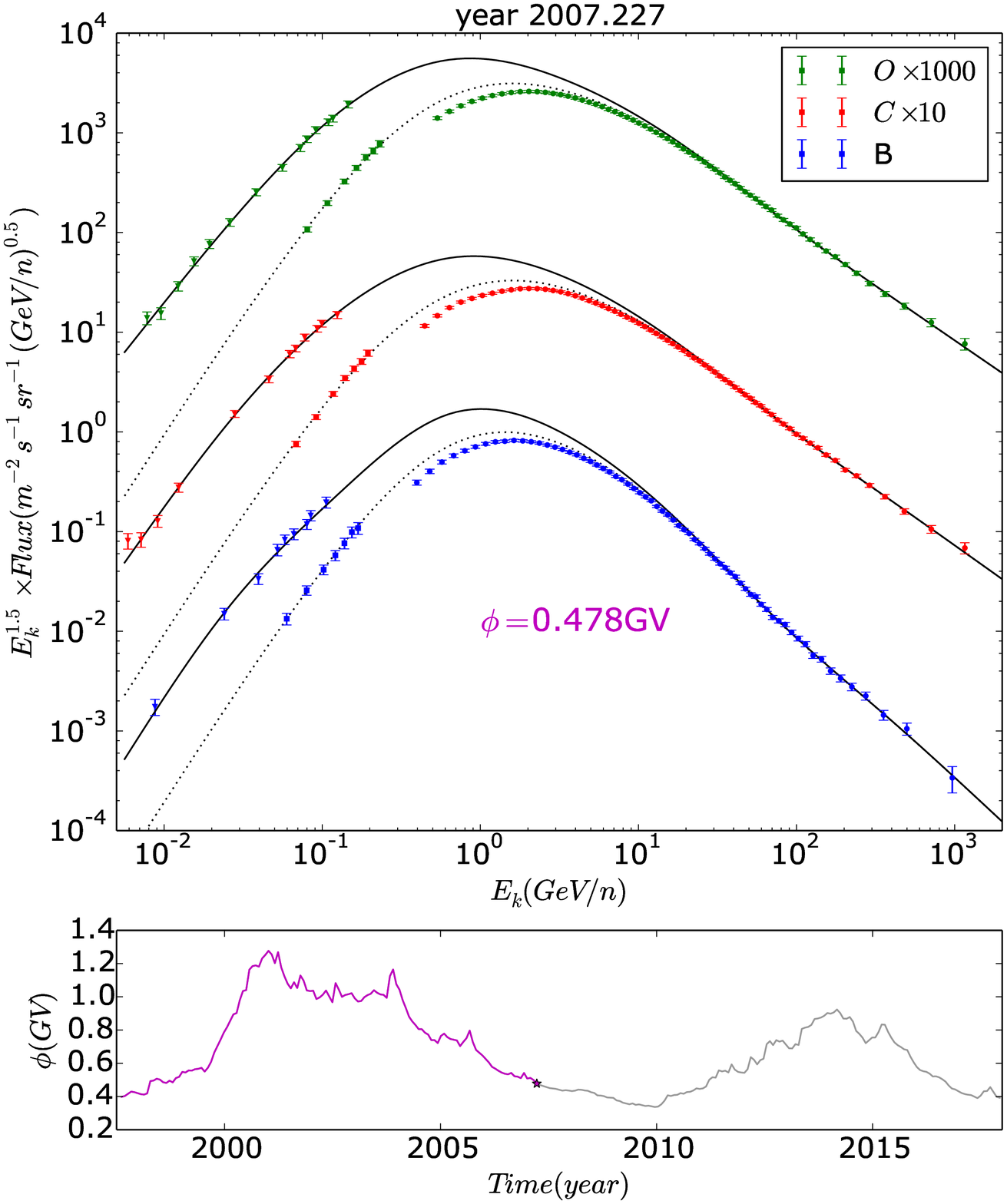}
\caption{Continuous of Fig~\ref{Fig.1}.}
\label{Fig.2}
\end{figure}

\begin{figure}[ht!]
\centering
\includegraphics[scale=0.3]{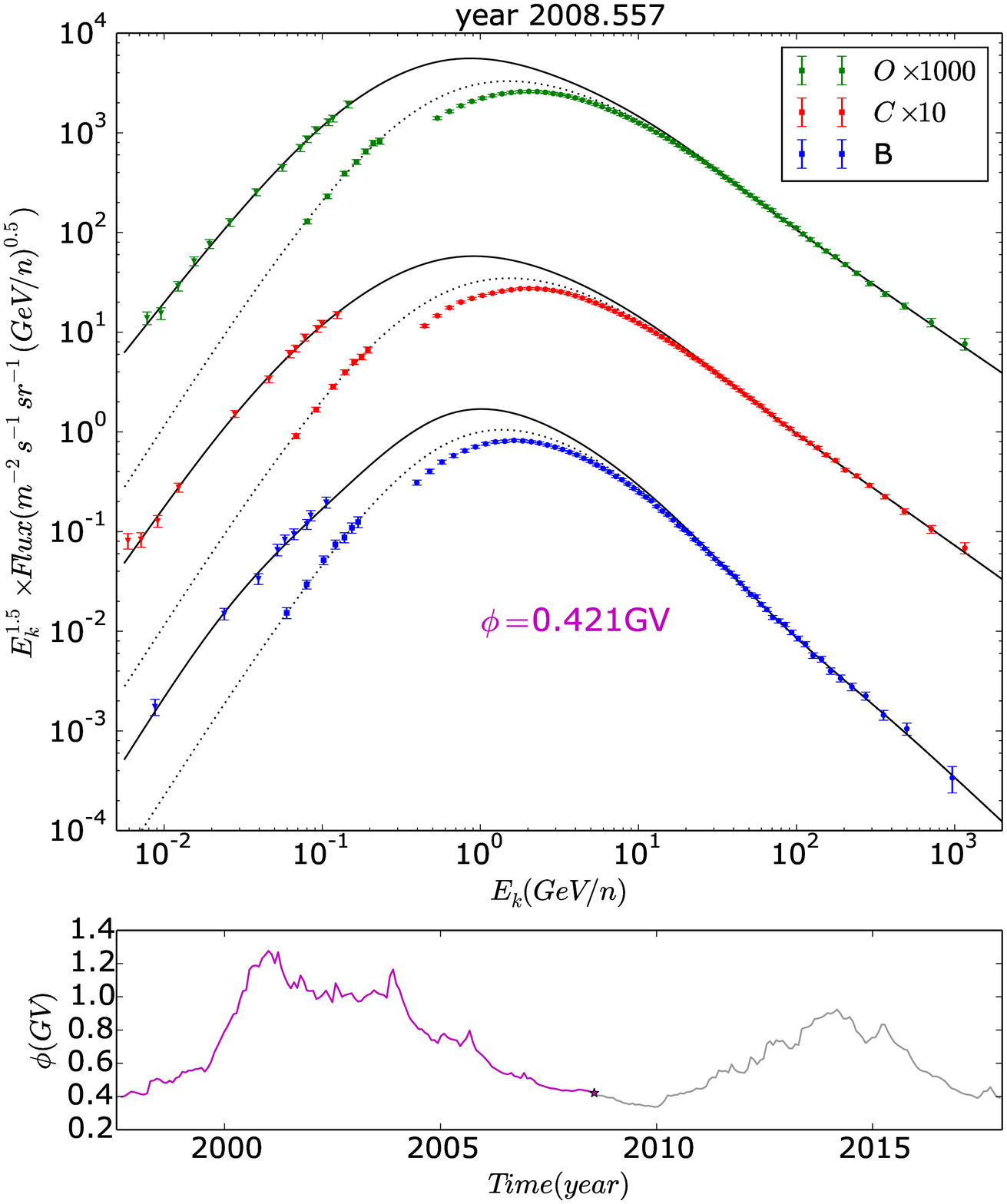}
\includegraphics[scale=0.3]{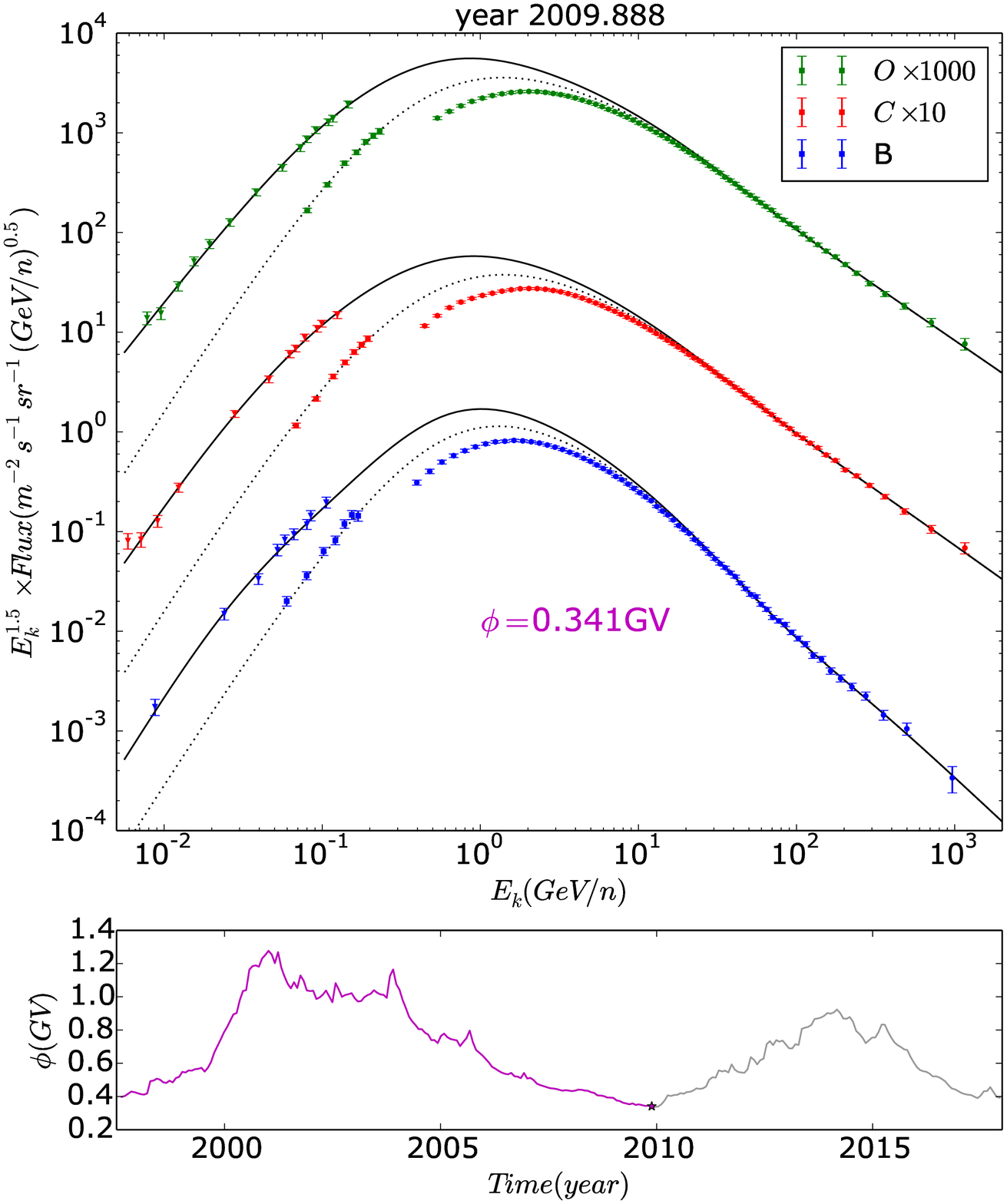}
\includegraphics[scale=0.3]{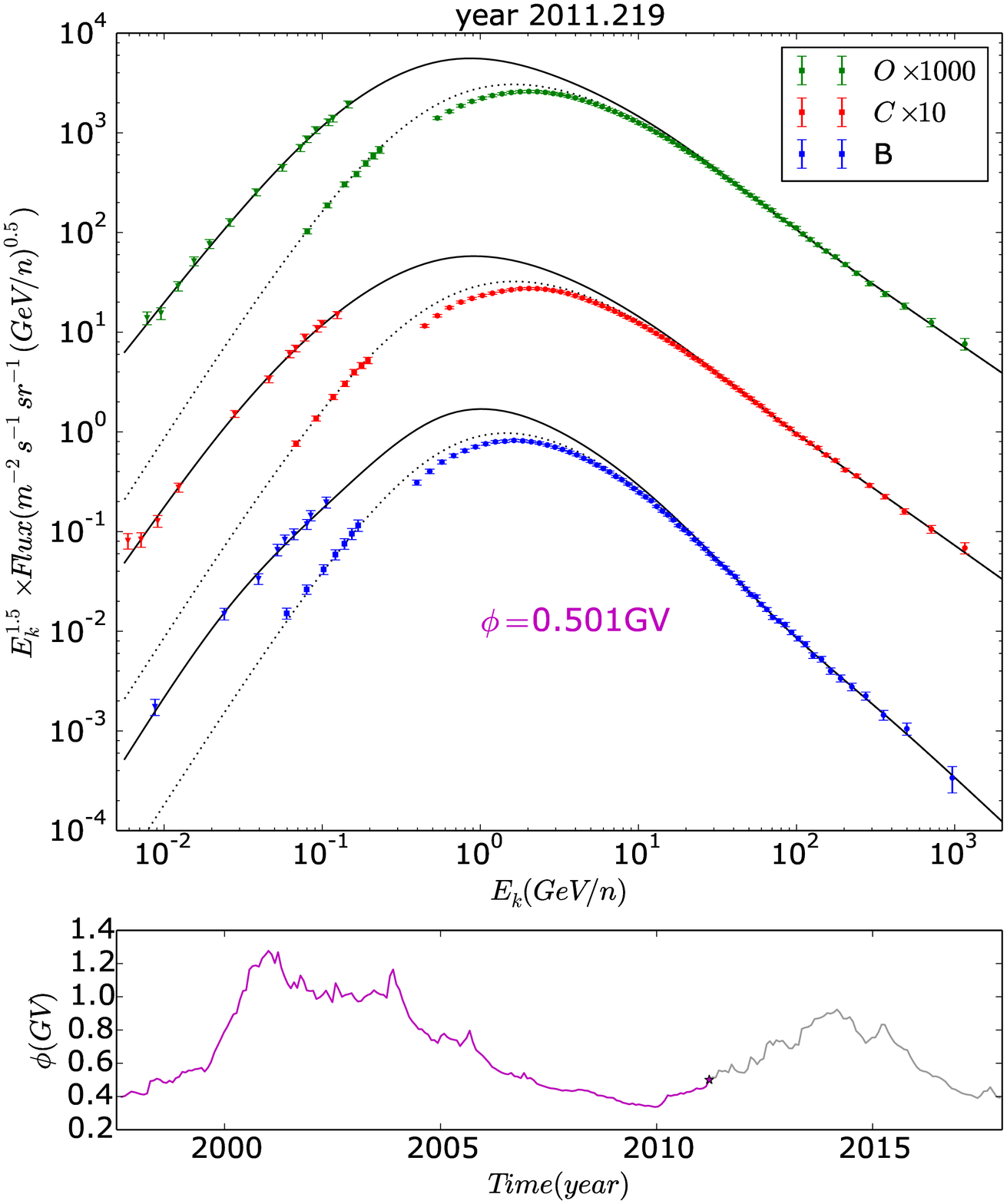}
\includegraphics[scale=0.3]{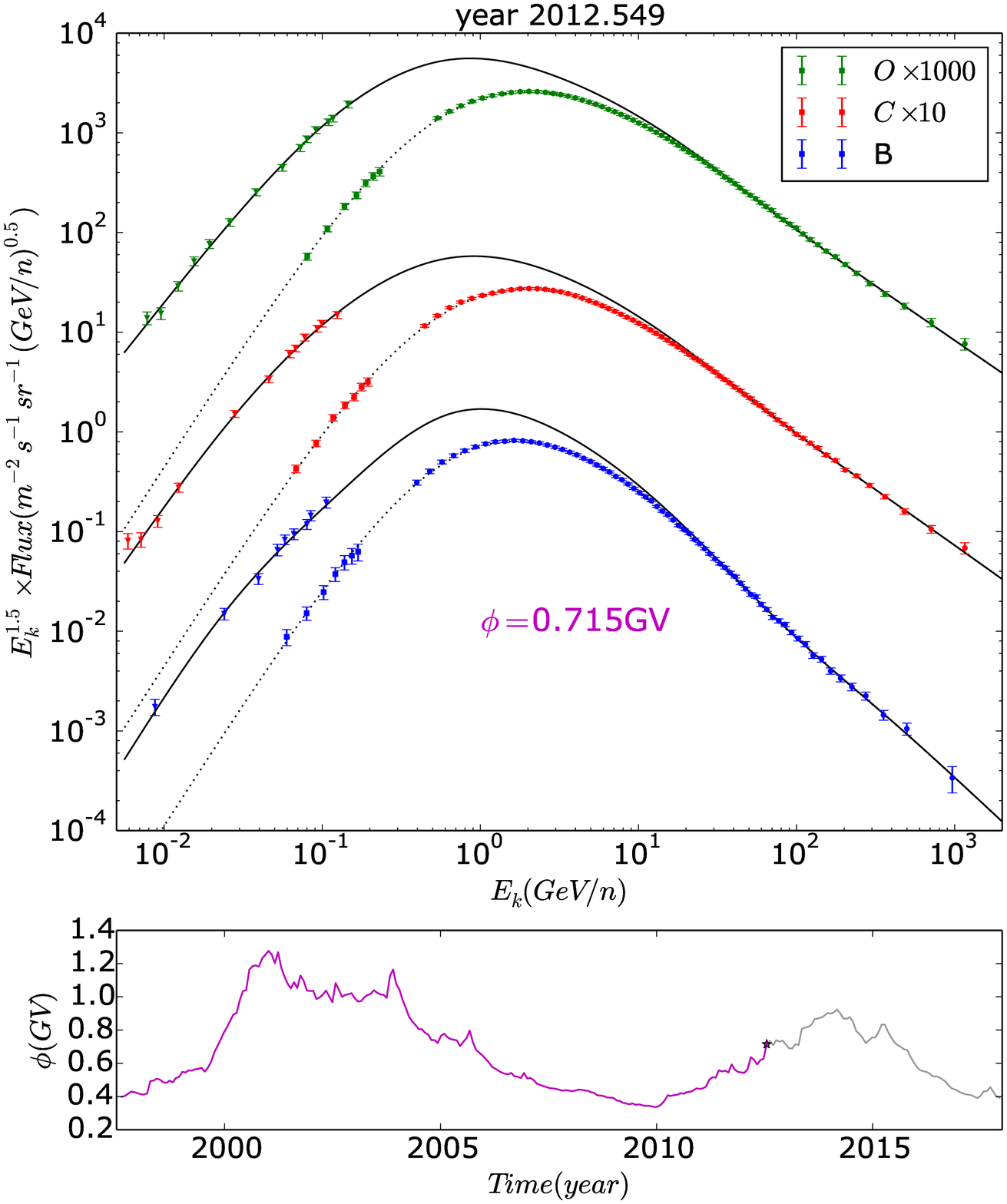}
\caption{Continuous of Fig~\ref{Fig.1}.}
\label{Fig.3}
\end{figure}

\begin{figure}[ht!]
\centering
\includegraphics[scale=0.3]{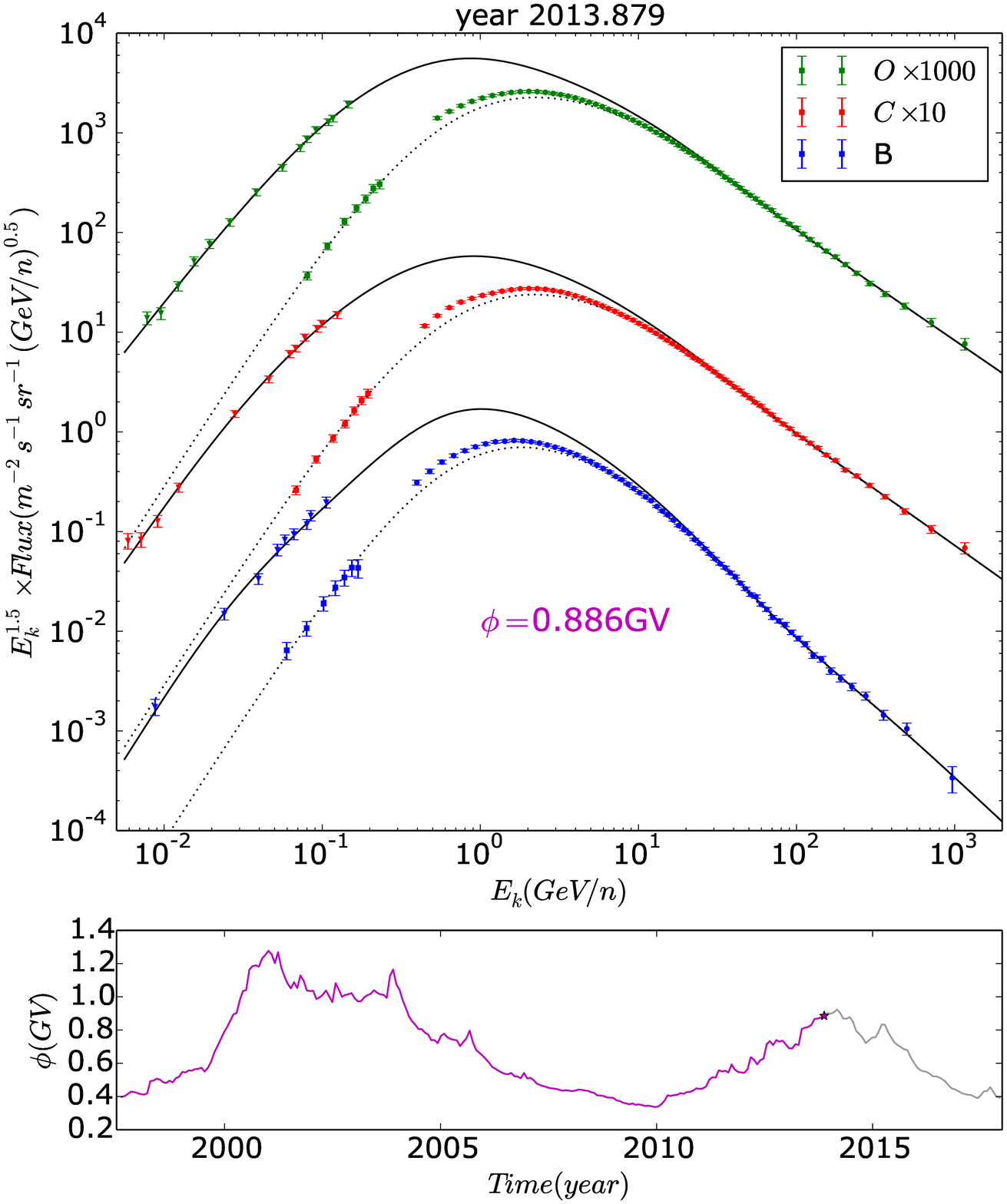}
\includegraphics[scale=0.3]{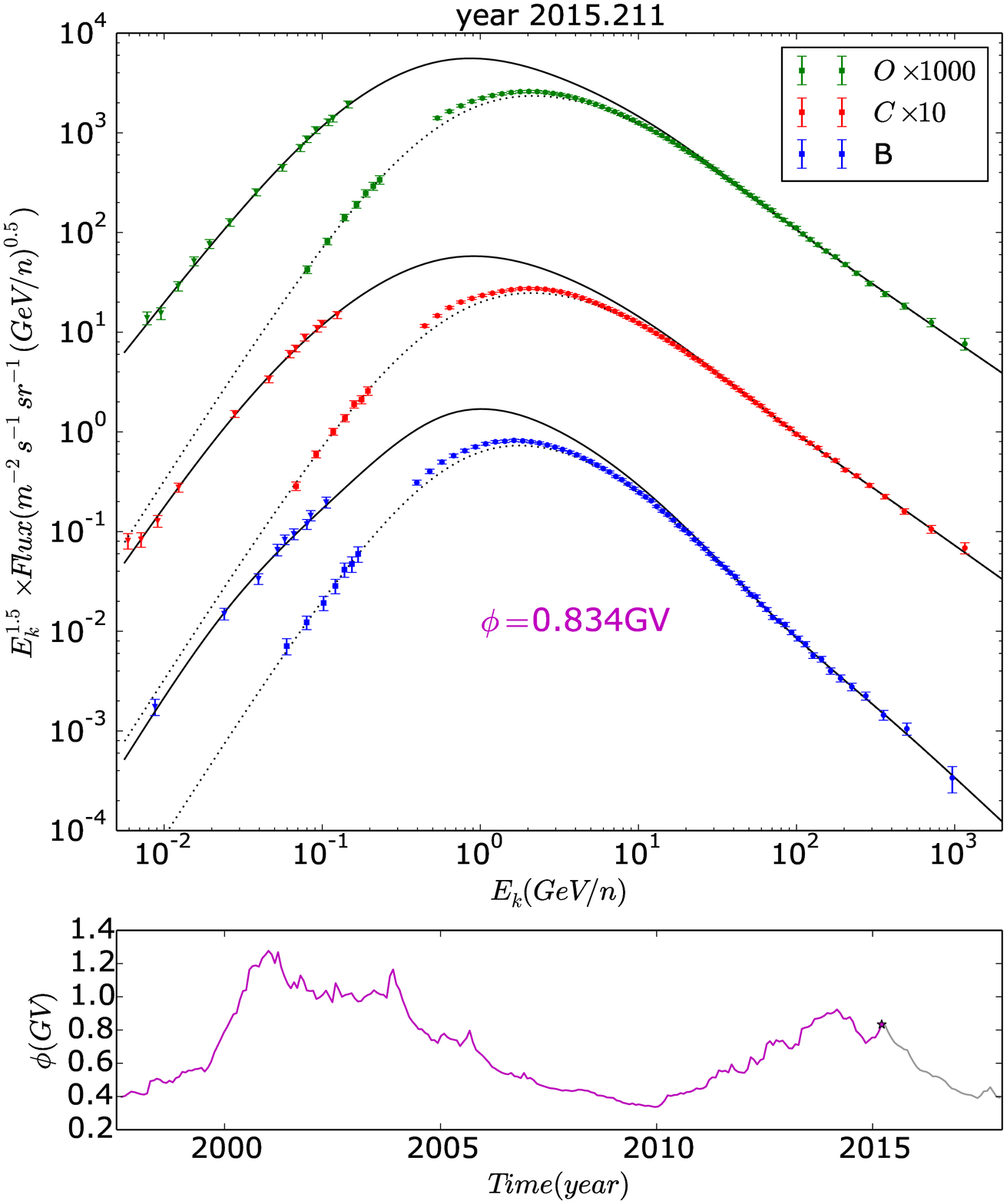}
\includegraphics[scale=0.3]{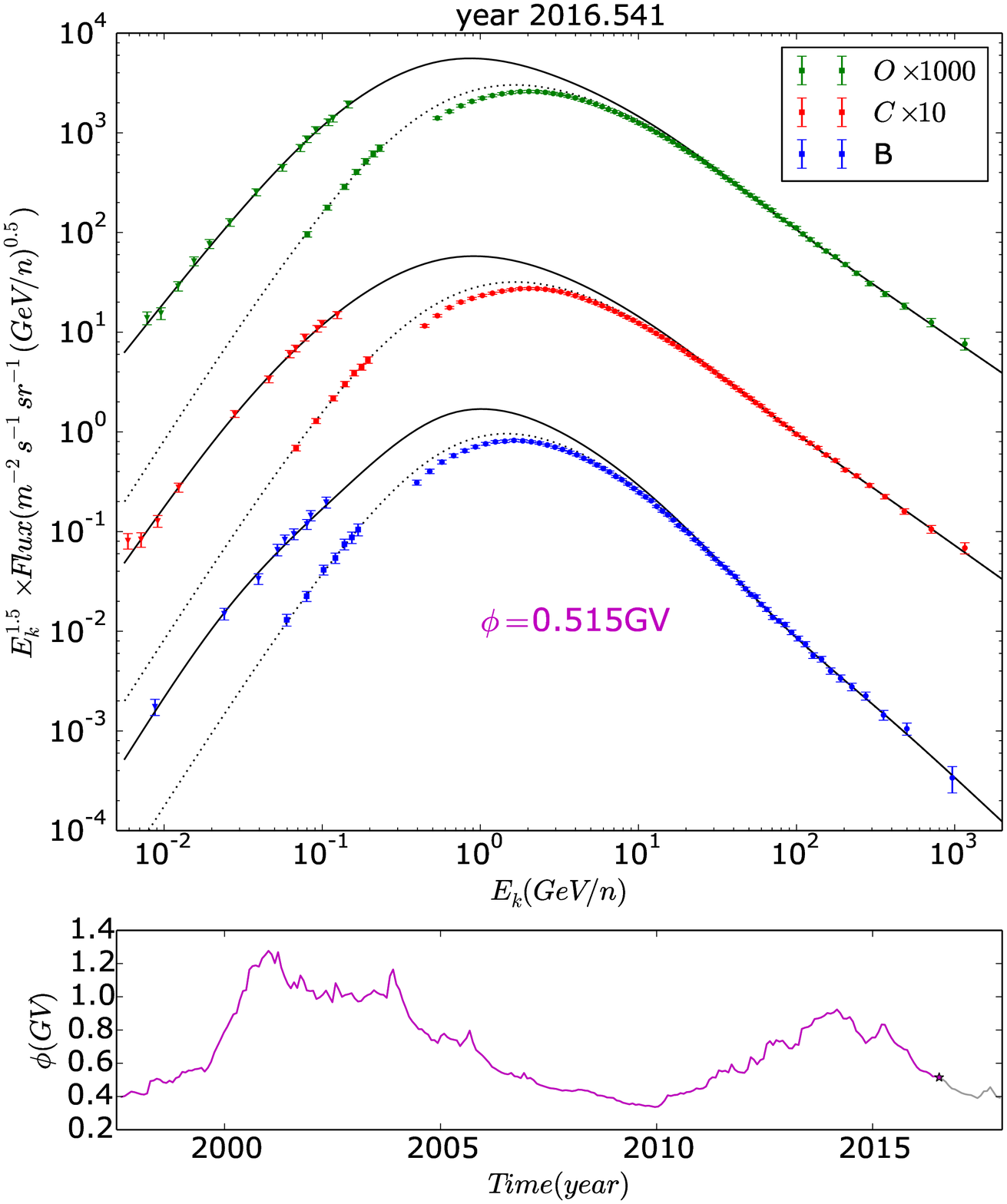}
\includegraphics[scale=0.3]{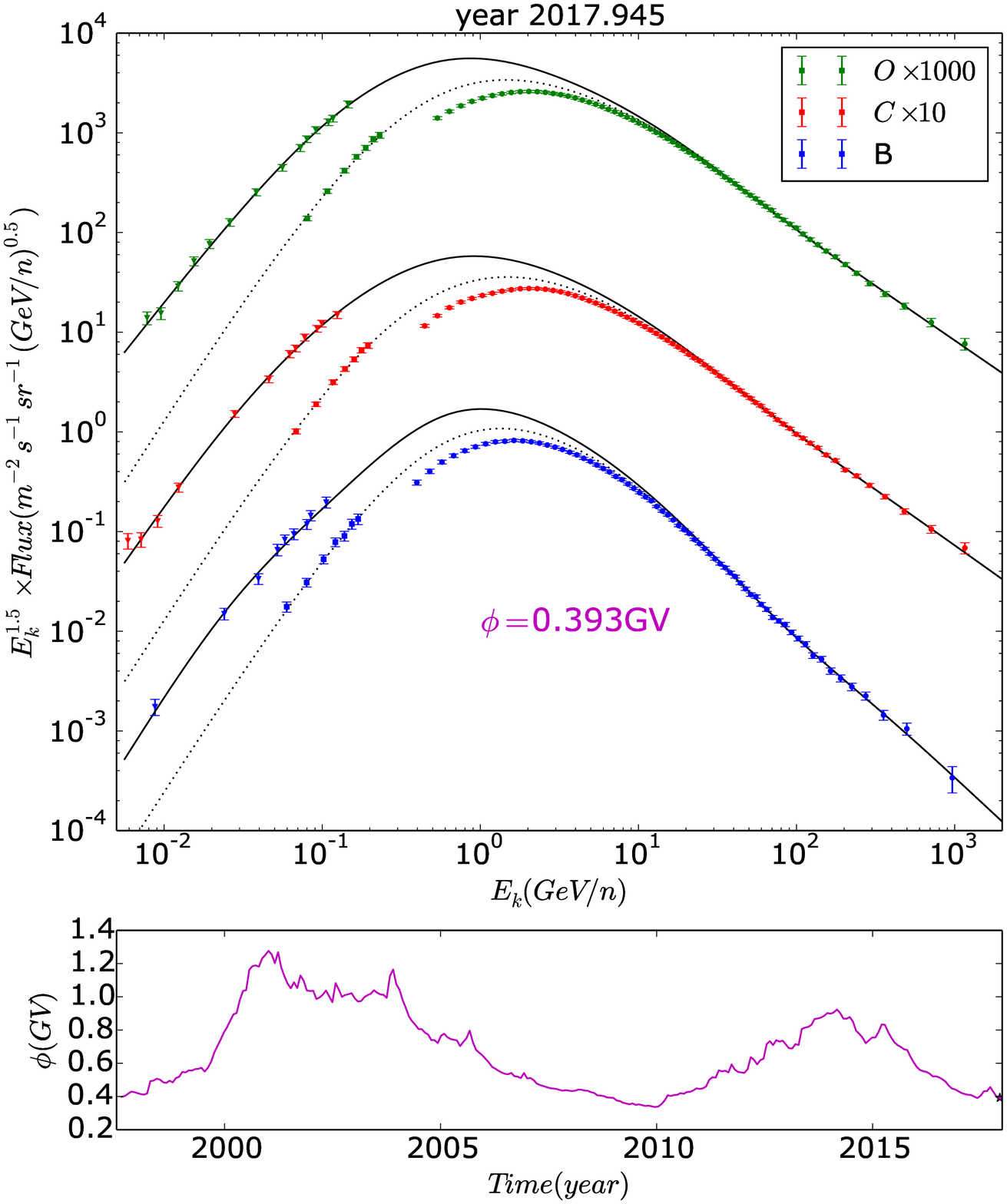}
\caption{Continuous of Fig~\ref{Fig.1}.}
\label{Fig.4}
\end{figure}

\end{document}